\documentclass[12pt]{article}
\usepackage{amsmath,amsfonts}
\makeatletter \@addtoreset{equation}{section}

\usepackage{cite}
\usepackage{bm}
\usepackage{dcolumn}
\makeatother
\def\one{{\hbox{ 1\kern-.8mm l}}}
\newcommand{\Dslash}{\not{\hbox{\kern-4pt $D$}}}
\newcommand{\pdslash}{\not{\hbox{\kern-2pt $\partial$}}}
\newcommand{\be}{\begin{equation}}
\newcommand{\bea}{\begin{eqnarray}}
\newcommand{\eea}{\end{eqnarray}}
\newcommand{\ba}{\begin{array}}
\newcommand{\ea}{\end{array}}
\newcommand{\ee}{\end{equation}}

\newcommand{\bss}{\begin{split}}
\newcommand{\ess}{\end{split}}

\begin{document}

\begin{titlepage}
\vspace*{1mm}%
\hfill%
\vbox{
    \halign{#\hfil        \cr
           IPM/P-2010/020 \cr
                     } 
      }  
\vspace*{15mm}%
\begin{center}

{{\Large {\bf  Holographic Renormalization of New Massive Gravity}}}

\vspace*{15mm} \vspace*{1mm} {Mohsen Alishahiha$^{a}$ and Ali
Naseh$^{a,b}$}

 \vspace*{1cm}

{\it ${}^a$ School of physics, Institute for Research in Fundamental Sciences (IPM)\\
P.O. Box 19395-5531, Tehran, Iran \\ }

\vspace*{.4cm}

{\it ${}^b$ Department of Physics, Sharif University of Technology \\
P.O. Box 11365-9161, Tehran, Iran}

\vspace*{.4cm}

email: alishah@ipm.ir, and naseh@ipm.ir

\vspace*{2cm}
\end{center}

\begin{abstract}
We study holographic renormalization for three dimensional  new massive gravity (NMG).
By studying the general fall off conditions for the metric allowed by the model at infinity,
we show that at the  critical point where the central charges of the dual CFT are zero
it contains a leading logarithmic behavior. In the context of AdS/CFT correspondence it can
be identified as a source for an irrelevant operator in the dual CFT. The presence of the
logarithmic fall off may be interpreted as the fact that the dual CFT would be a LCFT.

\end{abstract}

\end{titlepage}



\section{Introduction}

Although Einstein gravity in three dimensions even with cosmological constant has no propagating modes,
adding higher derivative terms to the action given raise to non-trivial propagating degrees
of freedom. The most well-known three dimensional gravity with higher derivative terms is topologically
massive gravity (TMG) \cite{{Deser:1981wh},{Deser:1982vy}} whose action is given by the Einstein-Hilbert
action plus three dimensional
gravitational Chern-Simons term. This model has propagating massive graviton. The model admits
several vacua including $AdS_3$ solution.

It is generically expected that adding higher derivative terms to the action leads to an instability
due to the present of ghost-like modes. Nevertheless it was shown \cite{Li:2008dq} that TMG model is stable
above the $AdS_3$ vacuum at a critical value of the coefficient of the Chern-Sioms term where
the model would be chiral in the sense that all the left moving excitations of the theory are
pure gauge. Therefore we are  only left with the right moving excitations\footnote{For a rigorous
definition of chiral CFT see \cite{Balasubramanian:2009bg}.}.

Soon after it was shown in \cite{Grumiller:2008qz} that the linearized equations of motion
at the critical point have a solution which may be interpreted as a left moving
excitation\footnote{Whether at the critical point the model is really chiral has been further investigated in several
papers including \cite{{Carlip:2008jk},{Park:2008yy},{Grumiller:2008pr},{Carlip:2008eq},{Carlip:2008qh},
{Giribet:2008bw},{Blagojevic:2008bn},{Li:2008yz}}.}.
However it is worth mentioning that this solution which has the same asymptotic behavior as
AdS wave solution \cite{{AyonBeato:2004fq},{AyonBeato:2005qq}} does not obey Brown-Henneaux
boundary conditions \cite{Brown:1986nw}. Therefore if we restrict ourselves to solutions
which satisfy the Brown-Henneaux conditions one may still have chiral theory. On the other hand
if we relax the boundary conditions the theory will not be chiral and indeed it was
conjectured in \cite{Grumiller:2008qz} that the dual theory (in the sense of AdS/CFT correspondence \cite{Maldacena:1997re}) could be a logarithmic CFT (LCFT) \cite{Gurarie:1993}.

Holographic renormalization of TMG model has been studied in \cite{Skenderis:2009nt} where
the authors have also computed one and two point functions of the dual CFT. It was shown  that
the two point functions of the left mover sector at the critical point are those of a LCFT with zero central charge
(See also \cite{Grumiller:2009mw}).

Another three dimensional gravity with higher derivative terms has been introduced in \cite{Bergshoeff:2009hq}.
The corresponding action is given by
 \bea\label{action}
 S&=&\frac{1}{16\pi G}\int_{R} d^3x\sqrt{-G}\left[R-2
\lambda\right]\ -\frac{1}{m^{2}}\frac{1}{16\pi G}\int_{R}
d^3x\sqrt{-G}\left[R^{\mu \nu}R_{\mu \nu}-\frac{3}{8}R^{2}\right].
\eea
This model, known as new massive gravity (NMG), admits several vacua including $AdS_3$ vacuum. It is
believed that NMG model on an asymptotically locally $AdS_3$ geometry may have a dual CFT whose central charges
are given by \cite{{Bergshoeff:2009aq},{Liu:2009kc}}( see also \cite{{Bergshoeff:2009tb},{Liu:2009bk}})
\be
c_L=c_R=\frac{3l}{2G}\left(1-\frac{1}{2m^2l^2}\right).
\ee
At the critical value, $m^2l^2=\frac{1}{2}$, where the central charges are zero it has been
shown \cite{AyonBeato:2009yq} that the model admit a new vacuum solution
which is not asymptotically locally $AdS_3$. More precisely at the critical point one finds AdS
wave solution whose metric is given by \cite{AyonBeato:2009yq}
\be
ds^2=\frac{d\rho^2}{4\rho^2}+\frac{1}{\rho}\bigg(-2dudv- F(\rho) du^2\bigg),
\ee
where $F(\rho)=(k_1(u) \rho+k_2(u))\ln \rho$ with $k_1(u)$ and $k_2(u)$ being arbitrary functions of $u$.

Indeed this solution has to be compared with the AdS wave solutions in
TMG model \cite{{AyonBeato:2004fq},{AyonBeato:2005qq}}. Therefore following the
observation of \cite{Grumiller:2008qz} one may wonder that the dual theory would be a LCFT.
By making use of similarities between  NMG and TMG the authors of \cite{Grumiller:2009sn}
have computed two point functions of the model and shown that the dual theory is, indeed, a LCFT.

The aim of this paper is to further explore different features of the NMG model. In particular
we will study the holographic renormalization of  NMG model. More precisely we observe that using
the Fefferman-Graham coordinates for the metric for the solution which is asymptotically locally $AdS_3$,
we do not need any boundary terms to have a well posed variational principle at the
critical point. We note, however, that due to the fact that we have a new solution at the
critical point one needs to change the asymptotic behavior of the metric to accommodate the new solution.
Actually the equations of motion allow us
to have a wider class of the boundary condition for the metric as follows
\bea
g_{ij}=b_{(0) ij}\log(\rho)+g_{(0) ij}+\left(b_{(2)
ij}\log(\rho)+ g_{(2) ij}\right)\rho+\cdots\ .
\eea
Note that when $b_{(0) ij}\neq 0$ the metric is not asymptotically locally $AdS_3$. Nevertheless
following \cite{Skenderis:2009nt} for the sufficiently small $b_{(0) ij}$ one may consider this
term perturbatively. Indeed using the AdS/CFT rules, $b_{(0) ij}$ may be considered as a source
for an irrelevant operator in the dual CFT and thus for the small $b_{(0) ij}$ we could still
use the CFT description.
Therefore this boundary condition allows two sources for two operators in the boundary CFT. This might be
considered as a sign that the dual CFT would be a LCFT.

Using the linearized equations of motion we will find the regularized
on-shell action up to quadratic terms. The regularized action can
then be used to find the correction functions of the corresponding
operators. Our rigorous derivation of the correlation functions based on holographic renormalization
is compatible with the results of
\cite{Grumiller:2009sn}. Indeed our results confirm that both sectors of the dual theory are indeed
two copies of a LCFT with zero central charges.

The paper is organized as follows. In the next section we study the variational principle for the NMG model
where we will show that for the solution we are interested in there is no need to add Gibbons-Hawking
boundary terms. In section 3 we linearize the equations of motion and solve them perturbatively.
In section 4 using the linearized equations we will find the on-shell regularized quadratic action
which can be used to read two point functions using the AdS/CFT dictionary. The last
section is devoted to the conclusions and discussions. Due to the fact that the computations and
the expressions of the equations are very lengthy, the detail of the equations are presented in
the several appendices.

\section{Variational principle and New Massive Gravity}

In this section we will study the variational principle for
NMG model\footnote{The variational principle for the NMG model has
also been studied in \cite{Hohm:2010jc}.}. To proceed we note that, in general, when
we consider the variation of a gravitational action with respect
to the metric schematically we get the following form for the
variation of the action
\bea\label{RG} \delta S=\frac{1}{16\pi
G}\int_{M}d^{d+1}x \sqrt{-G}\;\bigg[(...)\delta
G_{\mu\nu}\bigg]+\frac{1}{16\pi G}\int_{\partial M}
d^dx\sqrt{-\gamma}\bigg[(...)\delta G_{\mu\nu}+(...)\delta
G_{\mu\nu,\sigma}\bigg],
\eea
where $G_{\mu\nu}$ is the bulk metric while $\gamma_{\mu\nu}$ is the metric on
the boundary $\partial M$. As usual setting the first term (the volume term) to
zero one finds the equations of motion whereas the boundary terms
is set to zero by a proper boundary condition. Indeed the
second term in the equation \eqref{RG} is set to zero by
imposing Dirichlet boundary condition at the boundary $\delta G_{\mu\nu}|_{\partial
M}=0$. On the other hand to have a well-posed variational
principle with the Dirichlet boundary condition one has to add a boundary term to
the action to remove the last term in \eqref{RG}.
In general it is difficult to this boundary term for a
generic gravitational action, though for the cosmological
Einstein-Hilbert action the corresponding term is known; Gibbons-Hawking term. Since in
NMG  we have higher derivative terms, a priori, it is not
clear how to make the variational principle well-posed with the Dirichlet boundary condition.
Nevertheless using a specific solution of the equations of
motion we will show how to overcome the problem.

Starting with the action of NMG model \eqref{action}
one has
\bea\label{AA}
\delta S&=&\frac{1}{16\pi G}\int_{R}
d^3x \sqrt{-G}\;\left(R_{\mu\nu}-\frac{1}{2}G_{\mu\nu}R+\lambda
G_{\mu\nu}-\frac{1}{2m^{2}}K_{\mu\nu}\right)\;\delta G^{\mu\nu}  \cr & +&
\frac{1}{16\pi G}\int_{\partial R} d^2x \sqrt{-\gamma}\ n_{\mu}\bigg\{G^{\alpha\beta}\delta
\Gamma^{\mu}_{\alpha\beta}-G^{\alpha\mu}\delta\Gamma^{\beta}_{\alpha\beta}
-\frac{1}{m^2}\bigg[(2R^{\alpha\beta}-\frac{3}{4}R G^{\alpha\beta})\delta
\Gamma^{\mu}_{\alpha\beta}\cr
&&\;\;\;\;\;\;\;\;\;\;\;\;\;\;\;\;\;\;-(2R^{\alpha\mu}-\frac{3}{4}R
G^{\alpha\mu})\delta \Gamma^{\beta}_{\alpha\beta} -(2\nabla_{\beta}R^{\alpha\mu}G^{\beta\nu}-\nabla_{\beta}R^{\alpha\nu}G^{\mu\beta})\delta
G_{\alpha\nu}\bigg]\bigg\},
\eea
where
\bea\label{K}
K_{\mu\nu}&=&2\nabla^{2}R_{\mu\nu}-\frac{1}{2}(\nabla_{\mu}\nabla_{\nu}R+G_{\mu\nu}\nabla^{2}R)-8R_{\mu}^{\sigma}R_{\sigma\nu}
+\frac{9}{2}RR_{\mu\nu}\cr&+&G_{\mu\nu}\left(3R^{\alpha\beta}R_{\alpha\beta}-\frac{13}{8}R^{2}\right)
\eea
Setting the volume term to zero one finds
\bea\label{e.o.m}
R_{\mu\nu}-\frac{1}{2}G_{\mu\nu}R+\lambda
G_{\mu\nu}-\frac{1}{2m^{2}}K_{\mu\nu}=0,
\eea
which is the equations of motion whose trace is given by
\bea
R=-\frac{1}{m^{2}}\left(R^{\mu\nu}R_{\mu\nu}-\frac{3}{8}R^{2}\right)+6\lambda
\eea
It is easy to see that the model admits an AdS vacuum whose radius, $l$, is given
via the following expression
\bea
\lambda=-\frac{1}{l^{2}}(1+\frac{1}{4l^{2}m^{2}}).
\eea
To explore the validity of the variational principle, we
will consider a variation of the metric above a vacuum solution which is asymptotically locally $AdS_3$.
By making use of the Fefferman-Graham coordinates the metric of an asymptotically locally $AdS_3$
may be recast to the following form
\footnote{From now on we set $l=1$.}
\bea\label{F.G.e}
ds^{2}=\frac{d\rho^{2}}{4\rho^{2}}+\frac{1}{\rho}g_{ij}dx^{i}dx^{j}.
\eea
In this notation the boundary terms of the variation of the action \eqref{AA} read
\bea\label{var.act.bou.E.H}
 \delta S_{boundary}&=&\frac{1}{16\pi
G}\int_{\partial R} d^2x \sqrt{-\gamma}\ \bigg\{ B^{ij}\delta
g_{ij}+(2-\frac{1}{m^2})g^{ij}\delta g^{\prime}_{ij}-\frac{1}{m^2}{\tilde B}^{ij}g^{\prime}_{ij}\\
&&\;\;\;\;\;\;\;\;\;\;\;\;\;\;\;\;\;\;\;\;\;\;\;\;\;\;\;\;\;\;\;\;\; +\frac{1}{m^2}\bigg[\rho
g^{im}(\nabla^{n}g^{\prime}_{nm}-\nabla_{m}\;tr(g^{-1}g^{\prime}))g^{jk}\bigg]\delta
g_{kj,i}\bigg\}\nonumber
\eea
where
\bea
{\tilde B}^{ij}&=&\rho R(g)g^{ij}+2\rho
g^{ij}\;tr(g^{-1}g^{\prime})-4\rho^{2}(g^{-1}g^{\prime\prime}g^{-1})^{ij}
-2\rho^{2}(g^{-1}g^{\prime}g^{-1})^{ij}\;tr(g^{-1}g^{\prime})
\cr &+&
4\rho^2(g^{-1}[g^{\prime}g^{-1}g^{\prime}]g^{-1})^{ij}
-2\rho^{2}\;tr(g^{-1}g^{\prime\prime})g^{ij}+2\rho^{2}g^{ij}\;tr(g^{-1}g^{\prime}g^{-1}
g^{\prime}),
\eea
and the expression for $B^{ij}$ is given in the appendix A.

Taking into account that the boundary has no boundary the last term gives a new contribution to
the first term ( {\it i.e.} $B^{ij}$) which altogether do not contribute to the boundary due to the
fact that  we impose Dirichlet boundary condition at the boundary $\delta G_{\mu\nu}|_{\partial M}=0$.
On the other hand using the expression for the asymptotically locally $AdS_3$ solution,
\bea
g_{ij}=g_{(0) ij}+\left(b_{(2)ij}\log(\rho)+ g_{(2) ij}\right)\rho+\cdots,
\eea
one observes that ${\tilde B}^{ij}$ goes to zero as we approach the boundary. In other words this
term never reach the boundary. Therefore we are just left with the second term which can be
canceled by making use of the standard Gibbons-Hawking term with a proper coefficient.
In particular for $m^2=1/2$ this term is zero that means the model does not need
boundary term. Thus the variational principle is automatically well-posed for the model.

We note, however, that at the special value of $m^2=1/2$ the model admits a new vacuum solution
\cite{AyonBeato:2009yq} which is not asymptotically locally $AdS_3$. To accommodate this solution one
needs to change the asymptotic behavior of the metric as follows
\bea\label{asy.e}
g_{ij}=b_{(0) ij}\log(\rho)+g_{(0) ij}+\left(b_{(2)
ij}\log(\rho)+ g_{(2) ij}\right)\rho+\cdots\ .
\eea
It is important to note that with this new term ${\tilde B}^{ij}$ does not vanish as we
approach the boundary. It is worth mentioning that in the context of
the AdS/CFT correspondence
the new log term corresponds to a perturbation of the dual CFT with an irrelevant operator. Therefore
adding this term would destroy the conformal symmetry at UV.
Nevertheless following \cite{Skenderis:2009nt} we will assume that $b_{(0)ij}$ is
sufficiently small and this term can be treated perturbatively. Thus even for the case of $b_{(0)ij}\neq 0$,
we could still work for a solution which is asymptotically locally $AdS_3$.

As the conclusion for both $b_{(0) ij}=0$ and $b_{(0)ij}\neq 0$
cases the variational principle is well defined for NMG model without
adding any Gibbons-Hawking term for the special value $m^2=1/2$ which is the case
we will consider in this paper.

\section{Linearized analysis}

The main purpose of this article is to compute the correlation functions of
the energy momentum tensor of the CFT theory which is dual to the log gravity in
NMG model using the AdS/CFT correspondence. To do this one needs to find the on-shell
action which can be identified with the generating function for the energy momentum tensor
of the dual CFT. We note, however, that it is difficult to solve the equations of motion
above an $AdS$ vacuum in NMG model exactly. Nevertheless since in the present paper
we are only interested in two point functions it is enough to solve the equations
of motion at the linearized level. Note that in what follows we will also assume that
$b_{(0)ij}$ is sufficiently small such that we neglect higher powers of $b_{(0)ij}$.
In other words we will linearize the equations around an $AdS_3$ vacuum
whose metric is parametrized by
\bea\label{AdS.met}
d^{2}S=\frac{d^{2}\rho}{4\rho^{2}}+\frac{1}{\rho}\eta_{ij}dx^{i}dx^{j}.
\eea
With this notation we will consider the following perturbation
\bea\label{lin.met}
d^{2}S=\frac{d^{2}\rho}{4\rho^{2}}+\frac{1}{\rho}(\eta_{ij}+h_{ij})dx^{i}dx^{j}
\eea

\subsection{Linearized equations of motion}
In this subsection we would like to linearize the equations of motion in the case
of $m^2=1/2$
around the $AdS_3$ vacuum solution where we have $R=-6$.\footnote{See appendix C for arbitrary $m$.}
In other words we would like to study
space times which are asymptotically locally $AdS_3$. Taking into account that in this
case $\lambda=-3/2$ the equations of motion can be recast to the following form
\bea
E_{\mu\nu}\equiv R_{\mu\nu}+\frac{3}{2}G_{\mu\nu}-2\nabla^{2}R_{\mu\nu}+8R_{\mu}^{\rho}R_{\nu\rho}
+27R_{\mu\nu}-3G_{\mu\nu}(R^{\rho\sigma}R_{\rho\sigma})+\frac{117}{2}G_{\mu\nu}=0.
\eea
It is straightforward, though tedious to plug the metric \eqref{lin.met} into the above equations
of motion and keep only the linearized terms. Doing so one arrives at\footnote{${\tilde R}$ denotes the
linearized curvature.}
\bea\label{E.lin}
E_{\rho\rho}&=&4\rho^{2} tr(h^{\prime\prime\prime\prime})+16\rho  tr(h^{\prime\prime\prime})-12
 tr(h^{\prime\prime})+\rho \partial^{i}\partial_{i}
 tr(h^{\prime\prime})+\frac{4}{\rho} tr(h^{\prime})
 +2 \partial^{i}\partial_{i} tr(h^{\prime})
 \cr &&+\frac{6}{\rho^{2}}
 tr(h)+\frac{2}{\rho}\widetilde{R}(h)-2\partial^{i}\partial^{j}h^{\prime}_{ij}=0,\cr
E_{\rho i}&=&-4\rho^{2}\partial^{j}h^{\prime\prime\prime}_{ji}+4\rho^{2}\partial_{i}tr(h^{\prime\prime\prime})-8\rho\
\partial^{j}h^{\prime\prime}_{ji}+8\rho\
\partial_{i}tr(h^{\prime\prime})-\rho\
\partial^{n}\partial_{n}\partial^{j}h^{\prime}_{ji}-\partial_{i}\widetilde{R}(h)\cr &&
+\rho\ \partial^{j}\partial_{j}\partial_{i}\
tr(h^{\prime})-2\partial_{i}\
tr(h^{\prime})=0,\cr
E_{ij}&=&16\rho^{3}h^{\prime\prime\prime\prime}_{ij}+64\rho^{2}h^{\prime\prime\prime}_{ij}
-8\rho^{2}\eta_{ij}tr(h^{\prime\prime\prime})+32\rho
h^{\prime\prime}_{ij}-56\rho\ \eta_{ij}\
tr(h^{\prime\prime})-4\rho^{2}\eta_{ij}\widetilde{R}(h^{\prime\prime})\cr&&
+4\rho^{2}\partial^{m}\partial_{m}h^{\prime\prime}_{ij}+24\eta_{ij}\
tr(h^{\prime}) -2\rho\
\eta_{ij}\partial^{m}\partial_{m}tr(h^{\prime})+4\rho\
\partial_{i}\partial^{n}h^{\prime}_{nj}+4\rho\ \partial_{j}\partial^{n}h^{\prime}_{ni} \cr&&-8\rho\
\partial_{i}\partial{j}\ tr(h^{\prime})-8\rho\ \eta_{ij}\widetilde{R}(h^{\prime})
+12\eta_{ij}\widetilde{R}(h)-\rho\
\eta_{ij}\partial^{m}\partial_{m}\widetilde{R}(h) +\frac{24}{\rho}\
\eta_{ij}\ tr(h)=0.\cr &&
\eea
On the other hand the linearization of the trace condition, $R=-6$, leads to the following equation
\bea\label{tr.equ}
 -4\rho\ tr(h^{\prime\prime})+\widetilde{R}(h)+2\ tr(h^{\prime})=0.
\eea

\subsection{Near boundary solution of the linearized equations}

In this subsection we would like to solve the linearized equations of motion order by order around $\rho=0$
(near boundary). To proceed, motivated by the exact solutions of the NMG model we consider the following
expansion for the perturbation of the metric
\be\label{uu}
h_{ij}=b_{(0) ij}\log(\rho)+g_{(0)ij}+(b_{(2) ij}\log(\rho)+g_{(2) ij})\rho+\cdots\ .
\ee
As we have already mentioned in what follows we will also assume that $b_{(0)ij}$ is sufficiently small and therefore
the irrelevant deformation can be treated perturbatively.

From the above expansion one has
\bea\label{h.down}
h^{\prime}_{ij}&=&b_{(0) ij}\frac{1}{\rho}\ +b_{(2)ij}\log(\rho)+b_{(2) ij}+g_{(2) ij}+\cdots,\cr
h^{\prime\prime}_{ij}&=&-b_{(0) ij}\frac{1}{\rho^{2}}+b_{(2) ij}\frac{1}{\rho}+\cdots, \cr
h^{\prime\prime\prime}_{ij}&=&b_{(0) ij}\frac{2}{\rho^{3}}-b_{(2) ij}\frac{1}{\rho^{2}}+\cdots, \cr
h^{\prime\prime\prime\prime}_{ij}&=&-b_{(0) ij}\frac{6}{\rho^{4}}+b_{(2) ij}\frac{2}{\rho^{3}}+\cdots\ .
\eea
Plugging these expressions into the linearized equations of motion \eqref{E.lin} as well as the linearized trace equation \eqref{tr.equ} one can solve the resultant equations
order by order leading to algebraic equations for the parameters of the solutions. These equations can be solved
for the parameters $b_{(0) ij},g_{(0)ij},b_{(2) ij}$ and $g_{(2) ij}$.

To solve the equations we find that at orders $O(\frac{\log(\rho)}{\rho^{2}}), O(\frac{1}{\rho^{2}}),
O(\frac{\log(\rho)}{\rho})$ and $O(\frac{1}{\rho})$ non-trivial relations can be obtained only from
the $E_{\rho\rho}$ equation which are respectively given by
\be
 tr(b_{(0)})=0,\;\;\;\;\;tr(g_{(0)})=0,\;\;\;\;\;5tr(b_{(2)})+\widetilde{R}[b_{(0)}]=0,\;\;\;\;\;
6 tr(b_{(2)})-5 tr(g_{(2)})-\widetilde{R}[g_{(0)}]=0.
\ee
On the other hand at the order $O(\log(\rho))$ we get non-trivial relations from trace equation,
$E_{\rho i}$  and $E_{ij}$ as follows
\be
2tr(b_{(2)})+\widetilde{R}[b_{(0)}]=0,\;\;\;\;\partial_{i}[tr(b_{(2)})+\frac{1}{2}\widetilde{R}[b_{(0)}]]=0,
\;\;\;\;4tr(b_{(2)})+\widetilde{R}[b_{(0)}]=0.
\ee
These equations can be solved to find
\be
tr(b_{(0)})=tr(g_{(0)})=tr(b_{(2)})=\widetilde{R}[b_{(0)}]=0.
\ee
Using these results, at order $O(1)$ one finds
\bea
&{\rm from}&{\rm tr}\;\;\;:\;tr(g_{(2)})+\frac{1}{2}\widetilde{R}[g_{(0)}]=0,\cr
&{\rm from}&E_{\rho\rho}\;:\;\partial^{m}\partial^{n}b_{(2)mn}=0,\cr
&{\rm from}&E_{\rho i}\;:\;4\partial^{k}b_{(2)ki}+2\partial_{i}tr(g_{(2)})+\partial^{n}\partial_{n}
\partial^{k}b_{(0)ki}-\partial_{i}\widetilde{R}[g_{(0)}]=0,\cr
&{\rm from}&E_{ij}\;:\;
-12\eta_{ij}tr(g_{(2)})-3\eta_{ij}\widetilde{R}[g_{(0)}]
+\partial^{m}\partial_{m}b_{(0)ij}
+\partial_{i}\partial^{n}b_{(0)nj}-\partial_{j}\partial^{n}b_{(0)ni}=0,\cr &&
\eea
while at the order $O(\rho\log(\rho))$ we get
\bea
&{\rm from}&{\rm tr}\;\;\;:\;\widetilde{R}[b_{(2)}]=0,\cr
&{\rm from}&E_{\rho i}\;:\;
-\partial^{n}\partial_{n}\partial^{k}b_{(2)ki}-\partial_{i}\widetilde{R}[b_{(2)}]=0,\cr
&{\rm from}&E_{ij}\;:\;4\eta_{ij}\widetilde{R}[b_{(2)}]+4\partial_{i}\partial^{n}b_{(2)nj}
+4\partial_{j}\partial^{n}b_{(2)ni}=0.
\eea
Finally at the order $O(\rho)$ one has
\bea
&{\rm from}&{\rm tr}\;\;\;:\;\widetilde{R}[g_{(2)}]=0,\cr
&{\rm from}&E_{\rho i}\;:\;
\partial^{n}\partial_{n}[\partial_{i}tr(g_{(2)})-\partial^{k}b_{(2)ki}-\partial
^{k}g_{(2)ki}]-\partial_{i}\widetilde{R}[g_{(2)}]=0,\cr
&{\rm from}&E_{ij}\;:\;
-12\eta_{ij}\widetilde{R}[b_{(2)}]+4\eta_{ij}\widetilde{R}[g_{(2)}]-\eta_{ij}\partial^{m}\partial_{m}
\widetilde{R}[g_{(0)}]-2\eta_{ij}\partial^{m}\partial_{m}tr(g_{(2)})\cr
&&\;\;\;\;\;\;\;\;\;\;+4\partial^{m}\partial_{m}b_{(2)ij}+4\partial_{i}\partial^{m}
b_{(2)mj}+4\partial_{i}\partial^{m}g_{(2)mj}
+4\partial_{j}\partial^{m}b_{(2)mi}+4\partial_{j}\partial^{m}g_{(2)mi}=0.\cr &&
\eea
These equations can be solved leading to
\be
tr(g_{(2)})=0=\widetilde{R}[g_{(0)}]=\widetilde{R}[b_{(2)}]=\widetilde{R}[g_{(2)}]=0,
\ee
and
\bea
&&\partial^{m}[b_{(2)mi}+\frac{1}{4}\partial^{n}\partial_{n}b_{(0)mi}]=0,\;\;\;\;\partial^{m}\partial^{n}b_{(2)mn}=0,
\;\;\;\;\partial^{m}\partial_{m}\partial^{n}b_{(2)ni}=0,\cr
&&\partial^{m}\partial_{m}\partial^{n}g_{(2)ni}=0,\;\;\;\;
\partial_{i}\partial^{m}b_{(2)mj}+\partial_{j}\partial^{m}b_{(2)mi}=0,\cr
&&\partial^{m}\partial_{m}b_{(0)ij}=\partial_{i}\partial^{n}b_{(0)nj}+\partial_{j}\partial^{n}b_{(0)ni},\cr
&&\partial^{m}\partial_{m}b_{(2)ij}=-\partial_{i}\partial^{n}b_{(2)nj}-\partial_{j}\partial^{n}b_{(2)ni}
-\partial_{i}\partial^{n}g_{(2)nj}-\partial_{j}\partial^{n}g_{(2)ni}.
\eea

Note that from the near boundary solution one can not fix the metric completely. Indeed  to find a solution
of the linearized equations of motion one has to solve them exactly, as we will do in section 4.2.

\section{Correlation functions}

In this section we will compute two point functions of the energy momentum tensor of the CFT which
is supposed to be dual to the three dimensional log gravity given by the AdS wave solution of
NMG model at the critical value $m^2=1/2$. To do so, following the AdS/CFT correspondence one needs
to identify the bulk on-shell action as the generating function of the CFT. On other hand since the asymptotic
behavior of the metric has two divergent terms, one may think of these divergent terms as the sources
for the operators in the dual CFT. Since we are only interested in two point functions it is enough
to know the on-shell action up to quadratic terms. Then by functional variation with respect to the
sources we get the corresponding correlation functions.

\subsection{On-shell action and counterterms}

In this subsection we will evaluate the on-shell action up to
quadratic terms. We note, however, that in general the on-shell
action is divergent and one needs to regularized it by adding a
proper counterterms. Actually taking into account that at the
critical value $m^2=1/2$ the Gibbons-Hawking
boundary term is not needed, it is enough to linearized the original action \eqref{action} up to
quadratic terms. Indeed it is straightforward to compute the
corresponding on-shell action which turns out to be\footnote{For
the detail of the derivation see the appendix D.}
\bea\label{second.action}
S_{(2)}&=&\frac{1}{32\pi
G}\int_{\partial R}d^2x \bigg[-16\rho
h^{\prime\prime}_{ij}h^{ij}+8\rho^{2}h^{\prime\prime
}_{ij}h^{\prime\;ij}-8\rho^{2}h^{\prime\prime\prime}_{ij}h^{ij}
-4\rho\eta^{im}(\partial^{j}\partial^{n}h^{\prime}_{mn})h_{ij}
\cr &&\;\;\;\;\;\;\;\;\;\;\;\;\;\;\;\;\;\;\;\;\;\;\;\;\;
+2\rho\eta^{nl}\eta_{ij}(\partial_{n}\partial^{m}h^{\prime}_{ml})h^{ij}\bigg].
\eea
Note that raising and lowering indices are done with $g_{(0)
ij}$, and the traces are also taken by $g_{(0)}$.

Using the explicit form of the perturbation, $h_{ij}$, taken
into account that
\bea\label{h.up}
h_{ij}=b_{(0)ij}\log(\rho)+g_{(0)ij}+b_{(2)ij}\rho\log(\rho)+g_{(2)ij}\rho
+\cdots,
\eea
one finds that the action have a logarithmic
divergence as follows
\bea S_{(2)}=\frac{1}{32\pi G}\int_{\partial
R} d^2x (16b_{(0)}^{ij}b_{(2) ij})\log(\rho)+\cdots.
\eea
Therefore one needs to add a counterterms to cancel this
divergence to make the action finite. Note that to maintain the
covariance form of the action the counterterm has to be written in
terms of  $h_{ij}$ and its derivative. It is easy to see that in our case the
corresponding counterterm is given by
\bea
S_{c.t}=\frac{1}{32\pi G}\int_{\partial R} d^2x\  (-8\rho
h^{\prime\ ij}h^{\prime}_{ij}),
\eea
which in terms of the extrinsic curvature may be recast to the following form
\bea
S_{c.t}=\frac{1}{32\pi G}\int_{\partial R} d^2x\sqrt{-\gamma}
\;(-8 K_{ij}[h]K^{ij}[h]).
\eea
Here $\gamma_{ij}=\frac{\eta_{ij}}{\rho}$ is the background induced metric. Note that in writing
the above expression we have utilized the fact that the the extrinsic curvature is given by
$K_{i}^{j}=-\delta_{i}^{j}+\rho h^{\prime j}_{i}$.

As a conclusion  the total action up to the quadratic terms is
\bea
S_{(2),tot}&=&\frac{1}{32\pi G}\int_{\partial R}d^2x
\bigg[-16\rho h^{\prime\prime}_{ij}h^{ij}+8\rho^{2}h^{\prime\prime
}_{ij}h^{\prime\;ij}-8\rho^{2}h^{\prime\prime\prime}_{ij}h^{ij}
-4\rho\eta^{im}(\partial^{j}\partial^{n}h^{\prime}_{mn})h_{ij}
\cr &&\;\;\;\;\;\;\;\;\;\;\;\;\;\;\;\;\;\;\;\;\;\;\;\;\;
+2\rho\eta^{nl}\eta_{ij}(\partial_{n}\partial^{m}h^{\prime}_{ml}h^{ij})-8\rho
h^{\prime}_{ij}h^{\prime ij}\bigg].
 \eea
 It is now straightforward
to vary the action with respect to the sources to find the
one point functions. Note that since the action contains
higher derivative terms, the boundary condition is given by the
metric and its first derivative. Therefore we should vary the
action with respect to the metric $h_{ij}$ as well as its
derivative $h^{\prime}_{ij}$
\bea\label{second.action.final}
\frac{\delta S_{2,tot}}{\delta
h^{ij}}&=&\frac{1}{16\pi G}\bigg[-8\rho
h^{\prime\prime}_{ij}-4\rho^{2}h^{\prime\prime\prime}_{ij}-2\rho\eta_{im}(\partial_{j}\partial_{n}
h^{\prime\;mn})
+\rho\eta^{nl}\eta_{ij}(\partial_{n}\partial^{m}h^{\prime}_{ml})\bigg],
 \cr  \frac{\delta S_{2,tot}}{\delta h^{\prime\;ij}}&=&\frac{1}{16\pi
G}\bigg[4\rho^{2}h^{\prime\prime}_{ij}-4\rho
h^{\prime}_{ij}-2\rho\eta_{im}\partial_{n}\partial_{j}h^{mn}-\rho\eta_{kj}\eta_{mn}\partial^{k}
\partial_{i}h^{mn}\bigg].
\eea
By making use of the explicit form of the perturbation one
arrives at
\bea \frac{\delta S_{2,tot}}{\delta
h^{ij}}&=&\frac{1}{16\pi
G}\bigg[-4b_{(2)ij}-2\eta_{im}\partial_{j}\partial_{n}b_{(0)}^{mn}+\eta^{nl}\eta_{ij}
(\partial_{n}\partial^{m}b_{(0)ml})\bigg], \cr  \frac{\delta
S_{2,tot}}{\delta h^{\prime\;ij}}&=&\frac{1}{16\pi G}\bigg[-4\rho
g_{(2)ij}-4
b_{(2)ij}\rho\log(\rho)-2\eta_{im}(\partial_{n}\partial_{j}b_{(0)}^{mn})\rho\log(\rho)
\cr && -2\rho\eta_{im}(\partial_{n}\partial_{j}g_{(0)}^{mn})
-\eta_{kj}\eta_{mn}(\partial^{k}\partial_{i}g_{(0)}^{mn})\rho
-\eta_{kj}\eta_{mn}(\partial^{k}\partial_{i}b_{(0)}^{mn})\rho\log(\rho)\bigg]
\eea
Therefore we find
\bea\label{vv}
\langle T_{ij}\rangle&=&\lim_{\rho\rightarrow0}\frac{4\pi}{\sqrt{-\eta}}\frac{\delta
S_{2,tot}}{\delta h^{ij}}=
\frac{1}{4G}\bigg[-4b_{(2)ij}-2\eta_{im}\partial_{j}\partial_{n}b_{(0)}^{mn}
+\eta^{nl}\eta_{ij}(\partial_{n}\partial^{m}b_{(0)ml})\bigg], \cr
\langle t_{ij}\rangle&=&\lim_{\rho\rightarrow
0}\bigg(\frac{4\pi}{\rho\sqrt{-g}}\frac{\delta I}{\delta
h^{\prime\;ij}}-\log(\rho)\frac{4\pi}{\sqrt{-\eta}}\frac{\delta
I}{\delta h^{ij}}\bigg)  \cr
&=&\frac{1}{4G}\bigg[-4g_{(2)ij}-2\eta_{im}(\partial_{n}\partial_{j}g_{(0)}^{mn})\bigg].
\eea

\subsection{Holomorphic correlation functions}

The goal of this subsection is to compute two point functions of the dual CFT. This can be done by
varying the one point functions with respect to the sources. Therefore we need to know the explicit
expressions of the one point functions in terms of the sources. The corresponding
expressions can be found by solving the linearized equations of motion exactly.
To do this it is useful to make the following
change of coordinates (see appendix E for definition of our notations)
\be z=x-t\;\;\;\;\;\;\;\;\; \bar{z}=x+t.
\ee
In this notation the linearized equations of motion, \eqref{E.lin}, read
\bea
&&E_{\rho\rho}\;=\;2\rho^{4}h^{\prime\prime\prime\prime}_{z\bar{z}}+8\rho^{3}
h^{\prime\prime\prime}_{z\bar{z}}-6\rho^{2}h^{\prime\prime}_{z\bar{z}}
+2\rho^{3}\partial\bar{\partial}h^{\prime\prime}_{z\bar{z}}+2\rho
h^{\prime}_{z\bar{z}}+2\rho^{2}\partial\bar{\partial}h^{\prime}_{z\bar{z}}\cr
&&\;\;\;\;\;\;\;\;\;\;+3h_{z\bar{z}}+\rho\partial^{2}h_{\bar{z}\;\bar{z}}
+\rho\bar{\partial}^{2}h_{zz}-2\rho\partial\bar{\partial}
h_{z\bar{z}}-\rho^{2}\partial^{2}h^{\prime}_{\bar{z}\;\bar{z}}
-\rho^{2}\bar{\partial}^{2}h^{\prime}_{zz}=0,\cr &&\cr
&&E_{\rho1}=-2\rho^{2}\partial
h^{\prime\prime\prime}_{\bar{z}\;\bar{z}}
-2\rho^{2}\bar{\partial}h^{\prime\prime\prime}_{zz}
+2\rho^{2}\partial h^{\prime\prime\prime}_{z\bar{z}}
+2\rho^{2}\bar{\partial}h^{\prime\prime\prime}_{z\bar{z}}
-4\rho \partial h^{\prime\prime}_{\bar{z}\;\bar{z}}
-4\rho\bar{\partial}h^{\prime\prime}_{zz}\cr&&\;\;\;\;\;\;\;\;\;\;
+4\rho\partial h^{\prime\prime}_{z\bar{z}}+4\rho\bar{\partial
}h^{\prime\prime}_{z\bar{z}}-2\rho\partial^{2}\bar{\partial}h^{\prime}_{\bar{z}\;\bar{z}}
-2\rho\partial\bar{\partial}^{2}h^{\prime}_{zz}+2\rho\partial^{2}
\bar{\partial}h^{\prime}_{z\bar{z}}+2\rho\partial
\bar{\partial}^{2}h^{\prime}_{z\bar{z}}\cr&&\;\;\;\;\;\;\;\;\;\;
-\partial^{3}h_{\bar{z}\;\bar{z}}-\partial\bar{\partial}^{2}
h_{zz}+2\partial^{2}\bar{\partial}h_{z\bar{z}}-\partial^{2}
\bar{\partial}h_{\bar{z}\;\bar{z}}-\bar{\partial}^{3}
h_{zz}+2\partial\bar{\partial}^{2}h_{z\bar{z}}\cr&&\;\;\;\;\;\;\;\;\;\;
-2\partial h^{\prime}_{z\bar{z}}-2\bar{\partial}h^{\prime}_{z\bar{z}}=0,\cr &&\cr
&& E_{\rho2}=-2\rho^{2}\partial
h^{\prime\prime\prime}_{\bar{z}\;\bar{z}}
+2\rho^{2}\bar{\partial}h^{\prime\prime\prime}_{zz}
-2\rho^{2}\partial h^{\prime\prime\prime}_{z\bar{z}}
+2\rho^{2}\bar{\partial}h^{\prime\prime\prime}_{z\bar{z}}
-4\rho \partial h^{\prime\prime}_{\bar{z}\;\bar{z}}
+4\rho\bar{\partial}h^{\prime\prime}_{zz}\cr&&\;\;\;\;\;\;\;\;\;\;
-4\rho\partial h^{\prime\prime}_{z\bar{z}}+4\rho\bar{\partial }h^{\prime\prime}_{z\bar{z}}
-2\rho\partial^{2}\bar{\partial}h^{\prime}_{\bar{z}\;\bar{z}}
+2\rho\partial\bar{\partial}^{2}h^{\prime}_{zz}-2\rho\partial^{2}
\bar{\partial}h^{\prime}_{z\bar{z}}+2\rho\partial
\bar{\partial}^{2}h^{\prime}_{z\bar{z}}\cr&&\;\;\;\;\;\;\;\;\;\;
+\partial^{3}h_{\bar{z}\;\bar{z}}+\partial\bar{\partial}^{2}
h_{zz}-2\partial^{2}\bar{\partial}h_{z\bar{z}}-\partial^{2}
\bar{\partial}h_{\bar{z}\;\bar{z}}-\bar{\partial}^{3}
h_{zz}+2\partial\bar{\partial}^{2}h_{z\bar{z}}\cr&&\;\;\;\;\;\;\;\;\;\;
+2\partial h^{\prime}_{z\bar{z}}-2\bar{\partial}h^{\prime}_{z\bar{z}}=0,\cr &&\cr
&&E_{11}\;=\;\rho^{3}h^{\prime\prime\prime\prime}_{zz}+\rho^{3}h^{\prime
\prime\prime\prime}_{\bar{z}\;\bar{z}}+2\rho^{3}h^{\prime\prime
\prime\prime}_{z\bar{z}}+4\rho^{2}h^{\prime\prime\prime}_{zz}+
4\rho^{2}h^{\prime\prime\prime}_{\bar{z}\;\bar{z}}+6\rho^{2}
h^{3}_{z\bar{z}}\cr&&\;\;\;\;\;\;\;\;\;\;+2\rho
h^{\prime\prime} _{zz}+2\rho
h^{\prime\prime}_{\bar{z}\;\bar{z}}-10\rho
h^{\prime\prime}_{z\bar{z}}-\rho^{2}\partial^{2}
h^{\prime\prime}_{\bar{z}\;\bar{z}}-\rho^{2}\bar{\partial}^{2}
h^{\prime\prime}_{zz}+4\rho^{2}\partial\bar{\partial}
h^{\prime\prime}_{z\bar{z}}\cr&&\;\;\;\;\;\;\;\;\;\;+\rho^{2}\partial\bar{\partial}
h^{\prime\prime}_{zz}+\rho^{2}\partial\bar{\partial}
h^{\prime\prime}_{\bar{z}\;\bar{z}}+6h^{\prime}_{z\bar{z}}
-\rho \bar{\partial}^{2}h^{\prime}_{z\bar{z}}-\rho^{2}
\partial^{2}h^{\prime}_{z\bar{z}}-4\rho\partial\bar{\partial}
h^{\prime}_{z\bar{z}}\cr&&\;\;\;\;\;\;\;\;\;\;
-2\rho\partial^{2}h^{\prime}_{\bar{z}\;\bar{z}}-2\rho
\bar{\partial}^{2}h^{\prime}_{zz}+4\rho\partial\bar{\partial}
h^{\prime}_{z\bar{z}}+3\partial^{2}h_{\bar{z}\;\bar{z}}
+3\bar{\partial}^{2}h_{zz}-6\partial\bar{\partial}h_{z\bar{z}}
\cr&&\;\;\;\;\;\;\;\;\;\;-\rho\partial^{3}
\bar{\partial}h_{\bar{z}\;\bar{z}}-\rho\partial
\bar{\partial}^{3}h_{zz}+2\rho\partial^{2}\bar{\partial}^{2}
h_{z\bar{z}}+\frac{6}{\rho}h_{z\bar{z}}+\rho\partial^{2}
h^{\prime}_{zz}+\rho\bar{\partial}^{2}h^{\prime}_{\bar{z}\;
\bar{z}}\cr&&\;\;\;\;\;\;\;\;\;\;+\rho\partial\bar{\partial}
h^{\prime}_{zz}+\rho\partial\bar{\partial}h^{\prime}_{\bar{z}\;\bar{z}}=0,\cr &&\cr
&&E_{22}\;=\;\rho^{3}h^{\prime\prime\prime\prime}_{zz}+\rho^{3}h^{\prime
\prime\prime\prime}_{\bar{z}\;\bar{z}}-2\rho^{3}h^{\prime\prime
\prime\prime}_{z\bar{z}}+4\rho^{2}h^{\prime\prime\prime}_{zz}+
4\rho^{2}h^{\prime\prime\prime}_{\bar{z}\;\bar{z}}-6\rho^{2}
h^{3}_{z\bar{z}}\cr&&\;\;\;\;\;\;\;\;\;\;+2\rho
h^{\prime\prime} _{zz}+2\rho
h^{\prime\prime}_{\bar{z}\;\bar{z}}+10\rho
h^{\prime\prime}_{z\bar{z}}+\rho^{2}\partial^{2}
h^{\prime\prime}_{\bar{z}\;\bar{z}}+\rho^{2}\bar{\partial}^{2}
h^{\prime\prime}_{zz}-4\rho^{2}\partial\bar{\partial}
h^{\prime\prime}_{z\bar{z}}\cr&&\;\;\;\;\;\;\;\;\;\;+\rho^{2}\partial\bar{\partial}
h^{\prime\prime}_{zz}+\rho^{2}\partial\bar{\partial}
h^{\prime\prime}_{\bar{z}\;\bar{z}}-6h^{\prime}_{z\bar{z}}
-\rho \bar{\partial}^{2}h^{\prime}_{z\bar{z}}-\rho^{2}
\partial^{2}h^{\prime}_{z\bar{z}}+4\rho\partial\bar{\partial}
h^{\prime}_{z\bar{z}}\cr&&\;\;\;\;\;\;\;\;\;\;
+2\rho\partial^{2}h^{\prime}_{\bar{z}\;\bar{z}}+2\rho
\bar{\partial}^{2}h^{\prime}_{zz}-4\rho\partial\bar{\partial}
h^{\prime}_{z\bar{z}}-3\partial^{2}h_{\bar{z}\;\bar{z}}
-3\bar{\partial}^{2}h_{zz}+6\partial\bar{\partial}
h_{z\bar{z}}\cr&&\;\;\;\;\;\;\;\;\;\;+\rho\partial^{3}
\bar{\partial}h_{\bar{z}\;\bar{z}}+\rho\partial
\bar{\partial}^{3}h_{zz}-2\rho\partial^{2}\bar{\partial}^{2}
h_{z\bar{z}}-\frac{6}{\rho}h_{z\bar{z}}-\rho\partial^{2}
h^{\prime}_{zz}-\rho\bar{\partial}^{2}h^{\prime}_{\bar{z}\;
\bar{z}}\cr&&\;\;\;\;\;\;\;\;\;\;+\rho\partial\bar{\partial}
h^{\prime}_{zz}+\rho\partial\bar{\partial}
h^{\prime}_{\bar{z}\;\bar{z}}=0,\cr&&\cr
&&E_{12}\;=\;-\rho^{3}h^{\prime\prime\prime\prime}_{zz}+\rho^{3}
h^{\prime\prime\prime\prime}_{\bar{z}\;\bar{z}}-4\rho^{2}
h^{\prime\prime\prime}_{zz}+4\rho^{2}
h^{\prime\prime\prime}_{\bar{z}\;\bar{z}}-2\rho h^{\prime\prime}_{zz}+2\rho
h^{\prime\prime}_{\bar{z}\;\bar{z}}
-\rho^{2}\partial\bar{\partial}h^{2}_{zz}\cr&&\;\;\;\;\;\;\;\;\;\;
+\rho^{2}\partial\bar{\partial}h^{\prime\prime}_{\bar{z}\;\bar{z}}
+2\rho\partial^{2}h^{\prime}_{z\bar{z}}-2\rho
\bar{\partial}^{2}h^{\prime}_{z\bar{z}}-\rho\partial\bar{\partial}
h^{\prime}_{zz}+\rho\partial\bar{\partial}h^{\prime}_{\bar{z}\;\bar{z}}
-\rho\partial^{2}h^{\prime}_{z\bar{z}}\cr&&\;\;\;\;\;\;\;\;\;\;+\rho
\bar{\partial}^{2}h^{\prime}_{z\bar{z}}=0.
\eea
On the other hand from the linearized trace condition one finds
\bea
\partial^{2}h_{\bar{z}\;\bar{z}}+\bar{\partial}^{2}h_{zz}
-2\partial\bar{\partial}h_{z\bar{z}}-4\rho
h^{\prime\prime}_{z\bar{z}}
+2h^{\prime}_{z\bar{z}}=0.
\eea

By making use of the traceless conditions we have found in the section 3, the above equations may
be simplified. Indeed the above trace condition reads
\be\label{tr.z}
\partial^{2}h_{\bar{z}\;\bar{z}}+\bar{\partial}^{2}h_{zz}=0.
\ee
Moreover from the equations of motion we get
\bea
&&E_{\rho\rho}=\partial^{2}h^{\prime}_{\bar{z}\;\bar{z}}+\bar{\partial}^{2}h^{\prime}_{zz}=0,\cr&&\cr
&&E_{\rho1}=
2\rho^{2}(\partial h^{\prime\prime\prime}_{\bar{z}\;\bar{z}}+\bar{\partial}h^{\prime\prime\prime}_{zz})
+4\rho(\partial h^{\prime\prime}_{\bar{z}\;\bar{z}}+\bar{\partial}h^{\prime\prime}_{zz})
+2\rho(\partial^{2}\bar{\partial}h^{\prime}_{\bar{z}\;\bar{z}}+\partial\bar{\partial}^{2}h^{\prime}_{zz})\cr&&\;\;\;\;\;\;\;\;\;\;
+(\partial^{3}h_{\bar{z}\;\bar{z}}+\bar{\partial}^{3}h_{zz})
+(\partial\bar{\partial}^{2}
h_{zz}+\partial^{2}\bar{\partial}h_{\bar{z}\;\bar{z}})=0,\cr&&\cr
&& E_{\rho2}=2\rho^{2}(\partial
h^{\prime\prime\prime}_{\bar{z}\;\bar{z}} -\bar{\partial}h^{\prime\prime\prime}_{zz})
+4\rho(\partial h^{\prime\prime}_{\bar{z}\;\bar{z}}-\bar{\partial}h^{\prime\prime}_{zz})
+2\rho(\partial^{2}\bar{\partial}h^{\prime}_{\bar{z}\;\bar{z}}
-\partial\bar{\partial}^{2}h^{\prime}_{zz})\cr&&\;\;\;\;\;\;\;\;\;\;
-(\partial^{3}h_{\bar{z}\;\bar{z}}-\bar{\partial}^{3}h_{zz})-(\partial\bar{\partial}^{2}
h_{zz}-\partial^{2}\bar{\partial}h_{\bar{z}\;\bar{z}})
=0,\cr &&\cr
&&E_{11}=E_{22}=\rho^{2}(h^{\prime\prime\prime\prime}_{zz}+h^{\prime
\prime\prime\prime}_{\bar{z}\;\bar{z}})
+4\rho(h^{\prime\prime\prime}_{zz}+
h^{\prime\prime\prime}_{\bar{z}\;\bar{z}})
+2 (h^{\prime\prime}_{zz} +h^{\prime\prime}_{\bar{z}\;\bar{z}})\cr&&\;\;\;\;\;\;\;\;\;\;
+\rho(\partial\bar{\partial}
h^{\prime\prime}_{zz}+\partial\bar{\partial}
h^{\prime\prime}_{\bar{z}\;\bar{z}})
+(\partial\bar{\partial}
h^{\prime}_{zz}+\partial\bar{\partial}
h^{\prime}_{\bar{z}\;\bar{z}})=0, \cr&&\cr
&&E_{12}=\rho^{2}(h^{\prime\prime\prime\prime}_{zz}-
h^{\prime\prime\prime\prime}_{\bar{z}\;\bar{z}})
+4\rho(h^{\prime\prime\prime}_{zz}-
h^{\prime\prime\prime}_{\bar{z}\;\bar{z}})
+2( h^{\prime\prime}_{zz}-h^{\prime\prime}_{\bar{z}\;\bar{z}})
\cr&&\;\;\;\;\;\;\;\;\;\;
+\rho(\partial\bar{\partial}h^{\prime\prime}_{zz}
-\partial\bar{\partial}h^{\prime\prime}_{\bar{z}\;\bar{z}})
+(\partial\bar{\partial}h^{\prime}_{zz}-\partial\bar{\partial}h^{\prime}_{\bar{z}\;\bar{z}})=0.
\eea
Now the aim is to solve the above differential equations to find $h_{zz}$ and $h_{\bar{z}\bar{z}}$.
To proceed we will consider the following linear combinations of the above equations by which
we get decoupled equations for $h_{zz}$ and $h_{\bar{z}\bar{z}}$.
\bea
&&E_{11}+E_{12}=\rho^{2}h^{\prime\prime\prime\prime}_{\bar{z}\;\bar{z}}+4\rho
h^{3}_{\bar{z}\;\bar{z}}
+(-\alpha^{2}\rho+2)h^{\prime\prime}_{\bar{z}\;\bar{z}}
+\partial\bar{\partial}h^{\prime}_{\bar{z}\;\bar{z}}=0,\cr&&\cr
&&E_{11}-E_{12}==\rho^{2}h^{\prime\prime\prime\prime}_{zz}
+4\rho h^{3}_{zz}+(-\alpha^{2}\rho+2)
h^{\prime\prime}_{zz}+\partial\bar{\partial}h^{\prime}_{zz}=0
\eea
where $\alpha=\sqrt{-\partial\bar{\partial}}$. Note that in the momentum space, we have $\alpha\geq 0$.

The most general regular solution of the above equations
compatible with our boundary conditions are\footnote{ The second
solution of the equations is $I(0,2\alpha\sqrt{\rho})$ which is
not regular at $\rho\rightarrow \infty$.}
\bea
&&h_{zz}=c_{1}(z,\bar{z})+c_{2}(z,\bar{z})\ln(\rho)
+c_{3}(z,\bar{z})K(0,2\alpha\sqrt{\rho}),\cr&&\cr &&
h_{\bar{z}\;\bar{z}}=\bar{c}_{1}(z,\bar{z})
+\bar{c}_{2}(z,\bar{z})\ln(\rho)+\bar{c}_{3}(z,\bar{z})K(0,2\alpha\sqrt{\rho}).
\eea Using the asymptotic behavior of the Bessel function
$K(0,2\alpha\sqrt{\rho})$ at $\rho\rightarrow 0$, \bea
K(0,2\alpha\sqrt{\rho})=-\ln(\alpha\sqrt{\rho})-\gamma+[(1-\gamma)\alpha^{2}
-\ln(\alpha\sqrt{\rho})\alpha^{2}]\rho+... \;,
\eea
one finds\footnote{Note that this solution is consistent with $E_{\rho
1},E_{\rho 2 }$ equations if $\partial\bar{\partial}\bar{c}_2=0$ and $\partial^3\bar{\partial}^2\bar{c}_0=0$ .}
\bea
h_{zz}&=&[c_{2}(z,\bar{z})-\frac{1}{2}c_{3}(z,\bar{z})]\ln(\rho)
+[c_{1}(z,\bar{z})-(\gamma+\ln(\alpha))c_{3}(z,\bar{z})]\cr&&
-\frac{1}{2}[\alpha^{2}c_{3}(z,\bar{z})]\rho\ln(\rho)
+\bigg[[1-\gamma-\ln(\alpha)]\alpha^{2}c_{3}(z,\bar{z})\bigg]
\rho+..., \eea where $\gamma$ is the Euler-Mascheroni constant.
This has to be compared with \eqref{uu}. Doing so, one arrives at
\bea b_{(0)zz}&=&c_{2}(z,\bar{z})-\frac{1}{2}c_{3}(z,\bar{z}),\cr
g_{(0)zz}&=&c_{1}(z,\bar{z})-[\gamma+\ln(\alpha)]c_{3}(z,\bar{z}),\cr
b_{(2)zz}&=&-\frac{1}{2}[\alpha^{2}c_{3}(z,\bar{z})],\cr
g_{(2)zz}&=&[1-\gamma-\ln(\alpha)]\alpha^{2}c_{3}(z,\bar{z}).
\eea
It is useful to define
$c_{0}(z,\bar{z})=c_{2}(z,\bar{z})-\frac{1}{2}c_{3}(z,\bar{z})$.
In this notation the above equations can be recast to the
following form
\bea\label{h.zz}
b_{(0)zz}&=&c_{0}(z,\bar{z}),\cr
g_{(0)zz}&=&c_{1}(z,\bar{z})+2(\gamma+\ln(\alpha))(c_{0}(z,\bar{z})-
c_{2}(z,\bar{z})),\cr
b_{(2)zz}&=&\alpha^{2}(c_{0}(z,\bar{z})-c_{2}(z,\bar{z})),\cr
g_{(2)zz}&=&-2[1-\gamma-\ln(\alpha)]\alpha^{2}(c_{0}(z,\bar{z})-c_{2}(z,\bar{z})).
\eea Similarly one gets \bea\label{h./z/z}
b_{(0)\bar{z}\;\bar{z}}&=&\bar{c}_{0}(z,\bar{z}),\cr
g_{(0)\bar{z}\;\bar{z}}&=&\bar{c}_{1}(z,\bar{z})+2(\gamma+\ln(\alpha))(\bar{c}_{0}(z,\bar{z})-
\bar{c}_{2}(z,\bar{z})),\cr
b_{(2)\bar{z}\;\bar{z}}&=&\alpha^{2}(\bar{c}_{0}(z,\bar{z})-\bar{c}_{2}(z,\bar{z})),\cr
g_{(2)\bar{z}\;\bar{z}}&=&-2[1-\gamma-\ln(\alpha)]\alpha^{2}(\bar{c}_{0}(z,\bar{z})
-\bar{c}_{2}(z,\bar{z})).
\eea
On the other hand using the
equations (\ref{tr.z}),(\ref{h.zz}) and (\ref{h./z/z}) we find
\bea
c_{0}(z,\bar{z})&=&-\frac{\partial^{2}}{\bar{\partial}^{2}}
\;\bar{c}_{0}(z,\bar{z}),\;\;\;\;\;\;\;\;\;\;\;\;\;\;\;\;\;
c_{2}(z,\bar{z})=-\frac{\partial^{3}\bar{\partial}}{\partial
\bar{\partial}^{3}}\bar{c}_{2}(z,\bar{z}),\cr &&\cr
c_{1}(z,\bar{z})&=&-\frac{\partial^{2}}{\bar{\partial}^{2}}
\bar{c}_{1}(z,\bar{z})-2(\gamma+\ln(\alpha))\frac{\partial^{2}}{\bar{\partial}^{2}}
(\bar{c}_{0}(z,\bar{z})-\bar{c}_{2}(z,\bar{z}))\cr &&\cr
&&-2(\gamma+\ln(\alpha))(c_{0}(z,\bar{z})-c_{2}(z,\bar{z})).
\eea

Now we have all ingredients to evaluate two point functions. Note that since
NMG is a parity preserving model the both sectors of the dual CFT have the same structure.
Therefore it is enough to study only one sector of the dual CFT. Therefore in what follows we only consider the
{\it holomorphic} sector of the CFT.

In the $z,\bar{z}$ notation using the fact that
\bea
T_{zz}=\frac{1}{4}T_{11}+\frac{1}{4}T_{22} -\frac{1}{4}T_{12}
-\frac{1}{4}T_{21}.
\eea
from the equation \eqref{vv} one finds\footnote{See appendix  E.}
\bea\label{e.m.t.z}
\langle T_{zz}\rangle=\frac{1}{4G}[-4b_{(2)zz}+4\partial\bar{\partial}b_{(0)zz}],
\;\;\;\;\;\;\;
\langle t_{zz}\rangle=\frac{1}{4G}[-4g_{(2)zz}+4\partial\bar{\partial}g_{(0)zz}].
\eea
The two point functions can then be obtained from the general roles of AdS/CFT correspondence;
\bea
\langle T_{ij}...\rangle=i\frac{4\pi}{\sqrt{-g_{(0)}}}\frac{\delta}{\delta
g_{(0)}^{ij}}\langle...\rangle,\;\;\;\;\;\;\;
\langle t_{ij}\rangle=i\frac{4\pi}{\sqrt{-g_{(0)}}}\frac{\delta}{\delta
b_{(0)}^{ij}}\langle...\rangle .
\eea
On the other hand we note that
\bea
\frac{\delta}{\delta
b_{(0)}^{zz}}=-\frac{1}{4}\frac{\delta}{\delta
b_{(0)\bar{z}\;\bar{z}}}=-\frac{1}{4}\frac{\delta}{\delta
\bar{c}_{0}}.
\eea
Therefore
\bea
&&\frac{\delta}{\delta \bar{c}_{0}}b_{(2)zz}=-\frac{i}{2\pi}\frac{6}{z^{4}},\;\;\;\;\;\;\;\;\;
\frac{\delta}{\delta
\bar{c}_{0}}g_{(2)zz}=2[-1+\gamma+\ln(\alpha)]\frac{\partial^{3}}{\bar{\partial}}
\delta^{2}(z,\bar{z}),\cr &&\cr
&&\frac{\delta}{\delta
\bar{c}_{0}}[\partial\bar{\partial} b_{(0)zz}]=
\frac{i}{2\pi}\frac{6}{z^{4}},\;\;\;\;\;\;
\frac{\delta}{\delta
\bar{c}_{0}}[\partial\bar{\partial}
g_{(0)zz}]=-2(\gamma+\ln(\alpha))\frac{\partial^{3}}{\bar{\partial}}
\delta^{2}(z,\bar{z})
\eea
Finally we arrive at\footnote{We use this relations:
$\frac{1}{\partial\bar{\partial}}\delta^{2}(z,\bar{z})=\frac{i}{2\pi}
\ln(m^{2}|z|^{2})\;\;\;\;\;\;\;\;\ln(\alpha)\frac{1}{\partial\bar{\partial}}
\delta^{2}(z,\bar{z})=-\frac{i}{8\pi}\ln^{2}(m^{2}|z|^{2})$.}

\bea\label{two.point}
&&\langle T_{zz}T_{zz}\rangle=i\frac{4\pi}{\sqrt{-g_{(0)}}}\frac{\delta}{\delta
g_{(0)}^{zz}}\langle T_{zz}\rangle=0, \cr &&
\langle T_{zz}t_{zz}\rangle=i\frac{4\pi}{\sqrt{-g_{(0)}}}\frac{\delta}{\delta
b_{(0)}^{zz}}\langle T_{zz}\rangle=\frac{6/G}{z^{4}}, \cr &&
\langle t_{zz}t_{zz}\rangle=i\frac{4\pi}{\sqrt{-g_{(0)}}}\frac{\delta}{\delta
b_{(0)}^{zz}}\langle t_{zz}\rangle=\frac{1}{G}\frac{10+24\gamma-12\ln(m^{2}|z|^{2})}{z^{4}}.
\eea
which is the main goal of the present paper.  Note that the
non-logarithmic piece in the $\langle t_{zz}t_{zz}\rangle$ correlation can be removed by a
shift in $t$ given by $t_{zz}\rightarrow
t_{zz}-\frac{10+24\gamma}{12}T_{zz}$.

Similarly one can show
\bea\label{two.point2}
\langle T_{\bar{z}\bar{z}}T_{\bar{z}\bar{z}}\rangle==0,\;\;\;\;
\langle T_{\bar{z}\bar{z}}t_{\bar{z}\bar{z}}\rangle=\frac{6/G}{z^{4}},\;\;\;\;
\langle t_{\bar{z}\bar{z}}t_{\bar{z}\bar{z}}\rangle=-\frac{12}{G}\frac{\ln(m^{2}|z|^{2})}{z^{4}}.
\eea
The other correlations are zero up to a contact term.

\section{Conclusions}

In this paper we have explored the holographic renormalization for NMG model which is a three
dimensional parity preserving massive gravity. We have shown that  at the critical value
$l^2m^2=1/2$ we do not need any boundary terms to make the variational
principle of NMG model well-posed with Dirichlet boundary condition.

Using a proper fall off for the metric as a boundary condition we have computed the regularized
on shell action by adding a suitable counterterm. This in turn can be used to find correlation functions
of the dual field theory.

Indeed the main goal of this paper was to compute the two point function of the energy momentum tensor
of the dual CFT of NMG model at the critical point. Following the previous studies we would have
expected that the corresponding correlation function to have a LCFT-like behavior. Actually we have
found that the correlation functions, indeed, satisfy the expected expression for a LCFT.

Comparing our results \eqref{two.point} and \eqref{two.point2} with those in a
LCFT given by \cite{{Gurarie:1999yx},{MoghimiAraghi:2000qn}}
\bea\label{LCFT1}
&&\langle T_{zz}T_{zz}\rangle=\frac{c}{2z^4},
\cr && \langle T_{zz}t_{zz}\rangle=\frac{b}{2z^4}, \cr &&
\langle t_{zz}t_{zz}\rangle=-\frac{\log(z)[c\log(z)+2b]}{2z^4},
\eea
one observes that  the holographic two point functions have expected  from with central charge $c=\bar{c}=0$ and new anomaly $b=\bar{b}=\frac{12}{G}$. Having the same structure for left and right handed sectors with the equal
central charges and anomaly reflects the fact that the theory is parity preserving.

\section*{Acknowledgments}

We would like to thank H. Afshar, R. Fareghbal, D. Grumiller, A. Mosaffa and S. Rouhani for useful discussions.
Special thank to M. Taylor for carefully reading the paper and pointing out
a mistake in the early version of the manuscript.

\appendix

\section{Boundary term in the Fefferman-Graham coordinates }

\bea
B^{ij}&=&(-\frac{3} {\rho}g^{ij}+\rho g^{\prime ij})-\frac{1}{m^{2}}
\bigg[-\frac{1}{2}R(g)g^{ij}+4(g^{-1}g^{\prime\prime}g^{-1})^{ij}+3\rho(g^{-1}g^{\prime}g^{-1})^{ij}
tr(g^{-1}g^{\prime})  \cr &
-&\frac{3}{2\rho}g^{ij}-6\rho(g^{-1}[g^{\prime}g^{-1}g^{\prime}]g^{-1})^{ij}+6\rho
\;tr(g^{-1}g^{\prime\prime})^{ij}-\frac{5}{2}\rho\;tr(g^{-1}g^{\prime}g^{-1}g^{\prime})g^{ij}+\frac{17}{4}g^{\prime
ij}  \cr
&-&2\rho^{2}([g^{-1}g^{\prime}]g^{-1})^{ij}\;tr(g^{-1}g^{\prime\prime})+\rho^{2}([g^{-1}g^{\prime}]g^{-1})^{ij}\;tr(g^{-1}g^{\prime}g^{-1}g^{\prime})
+4(g^{-1}g^{\prime}g^{-1})^{ij}  \cr &-&\rho
(g^{-1}g^{\prime})^{ij}\;tr(g^{-1}g^{\prime})+2\rho(g^{-1}[g^{\prime}g^{-1}g^{\prime}])^{ij}
-\rho\;(g^{-1}g^{\prime})^{ij}R(g) -\rho(g^{-1}g^{\prime
})^{ij}\;tr(g^{-1}g^{\prime}) \cr &
+&2\rho^{2}(g^{-1}g^{\prime}g^{\prime\prime
-1})^{ij}-4\rho^{2}(g^{-1}g^{\prime}[g^{\prime}g^{-1}g^{\prime}])^{ij}-2\rho
g^{\prime ij}\;tr(g^{-1}g^{\prime})-\frac{1}{2}\rho
R^{\prime}(g)g^{ij}-\frac{1}{2}\rho R(g)(g^{-1}g^{\prime}g^{-1})
\cr &
-&\rho(g^{-1}g^{\prime}g^{-1})^{ij}\;tr(g^{-1}g^{\prime})+2\rho^{2}(g^{-1}g^{(3)}g^{-1})^{ij}
+\rho^{2}(g^{-1}g^{\prime\prime}g^{-1})^{ij}\;tr(g^{-1}g^{\prime})-2\rho(g^{-1}[g^{\prime}g^{-1}g^{\prime}])^{ij}
 \cr &
-&2\rho^{2}(g^{-1}[g^{\prime}g^{-1}g^{\prime}]g^{-1})^{ij}+\rho^{2}(g^{-1}g^{\prime}g^{-1})^{ij}\partial_{\rho}\;tr(g^{-1}g^{\prime})
-\rho g^{\prime ij}\;tr(g^{-1}g^{\prime})+2\rho
(g^{\prime}g^{-1}g^{\prime})^{ij} \cr &
+&2\rho^{2}(g^{-1}g^{\prime}g^{\prime
-1})^{ij}\;tr(g^{-1}g^{\prime})-\rho R(g)g^{\prime ij}
 \cr
&+&2\rho^{2}([g^{-1}g^{\prime}]g^{\prime\prime})^{ij}+4\rho^{2}(g^{\prime}g^{\prime\prime})^{ij}-\frac{1}{2}\rho
R(g)([g^{-1}g^{\prime}]g^{-1})^{ij}-4\rho^{2}(g^{\prime}[g^{-1}g^{\prime}g^{\prime}])^{ij}
\cr & +&2\rho^{2}(g^{\prime}g^{\prime})^{ij}\;tr(g^{-1}g^{\prime})
+\rho^{2}([g^{-1}g^{\prime}g^{\prime
-1}])^{ij}\;tr(g^{-1}g^{\prime})+2\rho
(\partial^{j}g^{in})(\nabla^{k}g^{\prime}_{kn}-\nabla_{n}
tr(g^{-1}g^{-1}g^{\prime}))  \cr &
-&\rho([g^{-1}g^{\prime}]g^{-1})^{ij}\;tr(g^{-1}g^{\prime})+\rho
g^{mn}(\nabla^{k}g^{\prime}_{kn}-\nabla_{n}\;tr(g^{-1}g^{\prime}))g^{ij}_{,m}\cr
&+&\rho
g^{mn}g^{kj}\Gamma^{i}_{mk}(\nabla^{l}g^{\prime}_{ln}-\nabla_{n}\;tr(g^{-1}g^{\prime}))+i\leftrightarrow
j\bigg]
 \eea

\section{Equations of motion in the Fefferman-Graham coordinates}

In this appendix we will work out how to write the equations of motion in terms of the
the Fefferman-Graham coordinates. To do this we need to express different component of the
equations of motion including $R, R_{\mu\nu}, R^{\mu\nu}R_{\mu\nu}, \nabla^2 R_{\mu\nu}$ in terms of
the Fefferman-Graham coordinates. It is straightforward, though tedious to show that
\bea
\label{R.S.F.G}
R&=&4\rho^{2}R_{\rho\rho}+\rho g^{ij}R_{ij}\cr &=&-6-4\rho^{2}
tr(g^{-1}g^{\prime\prime})+\rho^{2}
tr(g^{-1}g^{\prime}g^{-1}g^{\prime})+2\rho
tr(g^{-1}g^{\prime})-\rho^{2} tr^{2}(g^{-1}g^{\prime})\cr &+&2\rho^{2}
tr[g^{-1}(g^{\prime}g^{-1}g^{\prime})]+\rho R(g)
\eea

\bea
R_{\rho\rho}&=&-\frac{1}{2}tr(g^{-1}g^{\prime\prime})+\frac{1}{4}
tr(g^{-1}g^{\prime}g^{-1}g^{\prime})-\frac{1}{2\rho^2}\cr
R_{i\rho}&=&\frac{1}{2}[\nabla^{m}g^{\prime}_{mi}-\nabla_{i}tr(g^{-1}g^{\prime})]\cr
R_{ij}&=&\frac{1}{2}R(g)g_{ij}+g_{ij}tr(g^{-1}g^{\prime})-\rho[2g^{\prime\prime}_{ij}
+g^{\prime}_{ij}tr(g^{-1}g^{\prime})-2(g^{\prime}g^{-1}g^{\prime})_{ij}]-\frac{2}{\rho}g_{ij}
\eea
\bea
R_{\rho}^{\mu}R_{\mu\rho}&=&\rho^2[tr(g^{-1}g^{\prime\prime})]^{2}-\rho^{2}tr(g^{-1}g^{\prime\prime})
tr(g^{-1}g^{\prime}g^{-1}g^{\prime})+2 tr(g^{-1}g^{\prime\prime})-tr(g^{-1}g^{\prime}g^{-1}g^{\prime})+\frac{1}{\rho^{2}}\cr&
+&\frac{1}{4}\rho^{2}[tr(g^{-1}g^{\prime}g^{-1}g^{\prime})]^{2}+\frac{1}{4}\rho g^{ij}
[\nabla^{m}g^{\prime}_{mi}-\nabla_{i}tr(g^{-1}g^{\prime})][\nabla^{n}g^{\prime}_{nj}-\nabla_{j}tr(g^{-1}g^{\prime})]
\eea
\bea
R_{\rho}^{\mu}R_{\mu i}
&=&\big[-\rho^{2}tr(g^{-1}g^{\prime\prime})+\frac{1}{2}\rho^{2}tr(g^{-1}g^{\prime}g^{-1}g^{\prime})
-2+\frac{1}{4}\rho R(g)+\frac{1}{2}\rho tr(g^{-1}g^{\prime})\big] \cr &&\times \big[\nabla^{k}g^{\prime}_{ki}-\nabla_{i} tr(g^{-1}g^{\prime})\big]\cr &&
+\big[-\rho^{2}(g^{-1}g^{\prime\prime})-\frac{1}{2}\rho^{2}(g^{-1}g^{\prime}) tr(g^{-1}g^{\prime})+\rho^{2}
[g^{-1}(g^{\prime}g^{-1}g^{\prime})]\big]_{i}^{k} \cr &&
\big[\nabla^{n}g^{\prime}_{nk}-\nabla_{k} tr(g^{-1}g^{\prime})\big]
\eea
\bea
R_{i}^{\mu}R_{\mu j}&=&\frac{1}{4}\rho R^{2}(g)g_{ij}+\rho R(g)g_{ij} tr(g^{-1}g^{\prime})-2\rho^{2}R(g)g^{\prime\prime}_{ij}
-\rho^{2}R(g)g^{\prime}_{ij} tr(g^{-1}g^{\prime})+\frac{4}{\rho}g_{ij}\cr && +2\rho^{2}R(g)(g^{\prime}g^{-1}g^{\prime})_{ij}
-2R(g)\;g_{ij}+\rho\;g_{ij}[tr(g^{-1}g^{\prime})]^{2}-4\rho^{2}g^{\prime\prime}_{ij}\;tr(g^{-1}g^{\prime})\cr&&
-2\rho^{2}g^{\prime}_{ij}[tr(g^{-1}g^{\prime})]^{2}+4\rho^{2}(g^{\prime}g^{-1}g^{\prime})_{ij}tr(g^{-1}g^{\prime})
-4g_{ij}tr(g^{-1}g^{\prime})+4\rho^{3}[g^{-1}g^{\prime\prime}]_{i}^{k}g^{\prime\prime}_{kj} \cr &&
+4\rho^{3}[g^{-1}g^{\prime\prime}]_{i}^{k}g^{\prime}_{kj}tr(g^{-1}g^{\prime})-8\rho^{3}
[g^{-1}g^{\prime\prime}]_{i}^{k}(g^{\prime}g^{-1}g^{\prime})_{kj}+4\rho g^{\prime\prime}_{ij}
+4\rho[g^{-1}g^{\prime\prime}]_{i}^{k}g_{kj} \cr && +\rho^{3}[g^{-1}g^{\prime}]_{i}^{k}g^{\prime}_{kj}
[tr(g^{-1}g^{\prime})]^{2}-4\rho^{3}[g^{-1}g^{\prime}]_{i}^{k}(g^{\prime}g^{-1}g^{\prime})_{kj}
+4\rho g^{\prime}_{ij}tr(g^{-1}g^{\prime})  \cr && +4\rho^{3}[g^{-1}(g^{\prime}g^{-1}
g^{\prime})]_{i}^{k}(g^{\prime}g^{-1}g^{\prime})_{kj}-4\rho[g^{-1}(g^{\prime}g^{-1}g^{\prime})]_{i}^{k}
g_{kj}-4\rho(g^{\prime}g^{-1}g^{\prime})_{ij} \cr && +\rho^{2}[\nabla^{m}g^{\prime}_{mi}-\nabla_{i}
tr(g^{-1}g^{\prime})][\nabla^{n}g^{\prime}_{nj}-\nabla_{j}tr(g^{-1}g^{\prime})]
\eea
\bea
R^{\mu\nu}R_{\mu\nu}&=&4\rho^{4}[tr(g^{-1}g^{\prime\prime})]^{2}+\rho^{4}[tr(g^{-1}g^{\prime}g^{-1}g^{\prime})]^{2}
-4\rho^{4}tr(g^{-1}g^{\prime\prime})tr(g^{-1}g^{\prime}g^{-1}g^{\prime})\cr &&
-2\rho^{3}R(g)tr(g^{-1}g^{\prime\prime})-\rho^{3}R(g)[tr(g^{-1}g^{\prime})]^{2}+2\rho^{3}R(g)tr[g^{-1}(g^{\prime}g^{-1}
g^{\prime})]\cr && -4\rho R(g)+6\rho^{2}[tr(g^{-1}g^{\prime})]^{2}-4\rho^{3}tr(g^{-1}g^{\prime\prime})tr(g^{-1}g^{\prime})
-2\rho^{3}[tr(g^{-1}g^{\prime})]^{3}\cr &&+4\rho^{3}tr(g^{-1}g^{\prime})\;tr[g^{-1}(g^{\prime}g^{-1}g^{\prime})]-8\rho tr(
g^{-1}g^{\prime})+4\rho^{4}[g^{ij}g^{km}g^{\prime\prime}_{im}\;g^{\prime\prime}_{kj}] \cr &&+4\rho^{4}g^{ij}g^{km}
g^{\prime\prime}_{im}\;g^{\prime}_{kj}tr(g^{-1}g^{\prime})-8\rho^{4}g^{ij}g^{km}g^{\prime\prime}_{im}(g^{\prime}
g^{-1}g^{\prime})_{kj}+\frac{1}{2}\rho^{2}R^{2}(g)\cr && +\rho^{4}g^{ij}g^{km}g^{\prime}_{im}\;g^{\prime}_{kj}[tr(g^{-1}g^{\prime})]^{2}
-4\rho^{4}g^{ij}g^{km}g^{\prime}_{im}(g^{\prime}g^{-1}g^{\prime})_{jk}tr(g^{-1}g^{\prime})\cr &&+4\rho^{4}g^{ij}
g^{km}(g^{\prime}g^{-1}g^{\prime})_{im}(g^{\prime}g^{-1}g^{\prime})_{jk}-12\rho^{2}tr(g^{-1}g^{\prime}g^{-1}g^{\prime})
+16\rho^{2}tr(g^{-1}g^{\prime\prime}) \cr &&+12+2\rho^{2}R(g)tr(g^{-1}g^{\prime})+2\rho^{3}g^{ij}[\nabla^{k}g^{\prime}_{ki}-\nabla_{i}tr(g^{-1}g^{\prime})]
[\nabla^{m}g^{\prime}_{mj}-\nabla_{j}tr(g^{-1}g^{\prime})]\cr &&
\eea
\bea
\nabla^{2}R_{\rho\rho}&=&-2\rho^{2}\partial_{\rho}\partial_{\rho}tr(g^{-1}g^{\prime\prime})
+\rho^{2}\partial_{\rho}\partial_{\rho}tr(g^{-1}g^{\prime}g^{-1}g^{\prime})-10\rho \partial
_{\rho}tr(g^{-1}g^{\prime\prime})  \cr && +4\rho\partial_{\rho}tr(g^{-1}g^{\prime}g^{-1}g^{\prime})
+2tr(g^{-1}g^{\prime}g^{-1}g^{\prime})-3tr(g^{-1}g^{\prime\prime})-(g^{-1}g^{\prime})^{ik}
(g^{-1}g^{\prime})_{ik} \cr && -\frac{1}{2}\rho g^{ij}\partial_{i}\partial_{j}tr(g^{-1}g^{\prime\prime})
+\frac{1}{4}\rho g^{ij}\partial_{i}\partial_{j}tr(g^{-1}g^{\prime}g^{-1}g^{\prime})+\partial^{i}[
\nabla^{m}g^{\prime}_{mi}-\nabla_{i}tr(g^{-1}g^{\prime})] \cr &&-2\rho\partial_{\rho}
tr(g^{-1}g^{\prime\prime})-\rho^{2}tr(g^{-1}g^{\prime})\partial_{\rho}tr(g^{-1}g^{\prime\prime})+
\frac{1}{2}\rho^{2}tr(g^{-1}g^{\prime})\partial_{\rho}tr(g^{-1}g^{\prime}g^{-1}g^{\prime}) \cr &&
-4\rho tr(g^{-1}g^{\prime})tr(g^{-1}g^{\prime\prime})+2\rho tr(g^{-1}g^{\prime})tr(g^{-1}g^{\prime}g^{-1}g^{\prime})
+\frac{1}{2}\rho g^{ij}\Gamma_{ij}^{k}\partial_{k}tr(g^{-1}g^{\prime\prime}) \cr && -\frac{1}{4}\rho g^{ij}\Gamma
_{ij}^{k}\partial_{k}tr(g^{-1}g^{\prime}g^{-1}g^{\prime})-g^{ij}\Gamma_{ij}^{m}[\nabla^{n}g^{\prime}
_{nm}-\nabla_{m}tr(g^{-1}g^{\prime})]+\frac{1}{\rho}tr(g^{-1}g^{\prime})  \cr && +\frac{1}{2}\rho g^{ij}\Gamma_{ij}^{k}
(g^{-1}g^{\prime})_{k}^{m}[\nabla^{n}g^{\prime}_{nm}-\nabla_{m} tr(g^{-1}g^{\prime})]+\rho(g^{-1}g^{\prime})^{ij}
g^{\prime}_{ij}tr(g^{-1}g^{\prime})   \cr && +\rho^{2}[(g^{-1}g^{\prime})^{ik}
g^{\prime}_{ki}tr(g^{-1}g^{\prime\prime})]-\frac{1}{2}\rho^{2}[(g^{-1}g^{\prime})^{ik}g^{\prime}_{ki}
tr(g^{-1}g^{\prime}g^{-1}g^{\prime})]+(g^{-1}g^{\prime})^{ik}g^{\prime}_{ki} \cr &&+\frac{1}{2\rho}R(g)-\frac{1}{2}
R(g)tr(g^{-1}g^{\prime})-\frac{3}{2}[tr(g^{-1}g^{\prime})]^{2}+2\rho(g^{-1}g^{\prime})^{ij}g^{\prime\prime}_{ij}
 \cr && -2\rho(g^{-1}g^{\prime})_{i}^{j}
(g^{\prime}g^{-1}g^{\prime})_{j}^{i}+\frac{1}{4}\rho R(g)(g^{-1}g^{\prime})_{i}^{j}(g^{-1}g^{\prime})_{j}^{i}
+\frac{1}{2}\rho(g^{-1}g^{\prime})_{i}^{j}(g^{-1}g^{\prime})_{j}^{i}tr(g^{-1}g^{\prime})  \cr &&-\rho^{2}
(g^{-1}g^{\prime})^{ik}(g^{-1}g^{\prime})_{i}^{m}g^{\prime\prime}_{km}-\frac{1}{2}\rho (g^{-1}g^{\prime}
)^{ik}\partial_{i}[\nabla^{m}g^{\prime}_{mk}-\nabla_{k}tr(g^{-1}g^{\prime})] \cr && +\rho^{2}(g^{-1}g^{\prime})^{ik}
(g^{-1}g^{\prime})_{i}^{m}(g^{\prime}g^{-1}g^{\prime})_{km}-\frac{1}{2}\rho^{2}(g^{-1}g^{\prime})^{ik}
(g^{-1}g^{\prime})_{i}^{m}g^{\prime}_{km}tr(g^{-1}g^{\prime}) \cr && -\frac{1}{2}\rho g^{ij}\partial_{j}
[(g^{-1}g^{\prime})_{i}^{k}(\nabla^{m}g^{\prime}_{mk}-\nabla_{k}tr(g^{-1}g^{\prime}))]+\frac{1}{2}\rho(g^{-1}g^{\prime})^{ik}
\Gamma _{ik}^{m}[\nabla^{n}g^{\prime}_{nm}-\nabla_{m}tr(g^{-1}g^{\prime})]\cr &&
\eea
\bea
\nabla^{2}R_{i\rho}&=&\bigg[2\rho^{2}\partial_{\rho}\partial_{\rho}+\rho[g^{-1}
(g^{-1}g^{\prime})]^{kj}(g_{jk}-\rho g^{\prime}_{jk})+\frac{3}{2}\rho tr(
g^{-1}g^{\prime})+\rho^{2}tr(g^{-1}g^{\prime})\partial_{\rho} \cr && +5\rho\partial_{\rho}
+\frac{1}{2}\rho g^{jm}(\partial_{m}\partial_{j})\bigg]\bigg[\nabla^{k}g^{\prime}_{ki}
-\nabla_{i}tr(g^{-1}g^{\prime})\bigg] \cr && +\bigg[\rho\delta_{i}^{j}\partial_{\rho}
-\frac{1}{2}\rho^{2}(g^{-1}g^{\prime})_{i}^{m}(g^{-1}g^{\prime})_{m}^{j}-\frac{1}{2}\rho
g^{mk}(\partial_{m}\Gamma_{ik}^{j})-\frac{1}{2}\rho g^{mk}\Gamma_{ik}^{j}\partial_{m}
 \cr &&-\frac{1}{2}\rho tr(g^{-1}g^{\prime})\delta_{i}^{j}-\frac{1}{2}(g^{-1}g^{\prime}
)_{i}^{j}+\frac{1}{2}\rho(g^{-1}[g^{-1}g^{\prime}])^{mj}(g_{mi}-\rho g^{\prime}_{mi})
 \cr &&+\frac{1}{2}\rho^{2}tr(g^{-1}g^{\prime})(g^{-1}g^{\prime})_{i}^{j}-\delta_{i}^{j}
-\rho^{2}\partial_{\rho}(g^{-1}g^{\prime})_{i}^{j}\bigg]\bigg[\nabla^{k}g^{\prime}_{kj}
-\nabla_{j}tr(g^{-1}g^{\prime})\bigg] \cr && +\bigg[\rho g^{jk}\partial_{j}(g_{ik}+\rho
 g^{\prime}_{ik})+\rho g^{jk}(g_{ik}-\rho g^{\prime}_{ik})\partial_{j}+\rho\partial_{i}
 -\rho^{2}(g^{-1}g^{\prime})_{i}^{k}\partial_{k}\bigg] \cr &&\times \bigg[tr(g^{-1}g^
 {\prime\prime})-\frac{1}{2}tr(g^{-1}g^{\prime}g^{-1}g^{\prime})+\frac{1}{\rho^{2}}\bigg]
 \cr &&+\bigg[\frac{1}{2}g^{nk}\partial_{k}-\frac{1}{2}\rho g^{jk}\partial_{j}[-\frac{1}{\rho}
 \delta_{k}^{n}+(g^{-1}g^{\prime})_{k}^{n}]-\frac{1}{2}\rho  g^{jk}[-\frac{1}{\rho}\delta_{k}
 ^{n}+(g^{-1}g^{\prime})_{k}^{n}]\partial_{j} \cr &&-\frac{1}{2}g^{jk}\Gamma_{jk}^{n}+\frac{1}{2}
 g^{nk}\partial_{k}-\frac{1}{2}\rho[g^{-1}(g^{-1}g^{\prime})]^{kn}\partial_{k}\bigg]\bigg
 [\frac{1}{2}R(g)g_{in}+g_{in}tr(g^{-1}g^{\prime}) \cr &&
 -2\rho g^{\prime\prime}_{in}-\rho g^{\prime}_{in}tr(g^{-1}g^{\prime})+2\rho(g^{\prime}g^{-1}
 g^{\prime})_{in}-\frac{2}{\rho}g_{in}\bigg] \cr && -\frac{1}{2}g^{jk}\Gamma_{ik}^{n}\bigg[
 \frac{1}{2}R(g)g_{jn}+g_{jn}tr(g^{-1}g^{\prime})-2\rho g^{\prime\prime}_{jn}-\rho g^{\prime}_{jn}
 tr(g^{-1}g^{\prime})+2\rho(g^{\prime}g^{-1}g^{\prime})_{jn}\cr &&-\frac{2}{\rho}g_{jn}\bigg]
 -\rho g^{jk}\Gamma_{jk}^{m}\nabla_{m}R_{i\rho}-\rho g^{jk}\Gamma_{ij}^{m}\nabla_{k}R_{m\rho}
 +\frac{1}{2}\rho[g^{-1}(g^{-1}g^{\prime})]_{k}^{j}(\Gamma_{ik}^{n}R_{jm}+\Gamma_{jk}^{n}R_{in})\cr &&
\eea

\bea
\nabla^{2}R_{ij}&=&\bigg[2+4\rho^{2}\partial_{\rho}\partial_{\rho}+4\rho\partial_{\rho}
+\rho
g^{km}\partial_{m}\partial_{k}+(-4\rho+2\rho^{2}tr(g^{-1}g^{\prime}))\partial_{\rho}
-\rho g^{km}\Gamma_{km}^{n}\nabla_{n}\bigg]\cr&&
\bigg[\frac{1}{2}R(g)g_{ij} +g_{ij}tr(g^{-1}g^{\prime}) -2\rho
g^{\prime\prime}_{ij}-\rho
g^{\prime}_{ij}tr(g^{-1}g^{\prime})+2\rho(g^{\prime}g^{-1}g^{\prime})_{ij}
-\frac{2}{\rho}g_{ij}\bigg] \cr &&+\bigg[\big[4\rho\delta
_{i}^{n}\partial_{\rho}-4\rho^{2}(g^{-1}g^{\prime})_{i}^{n}\partial_{\rho}
-\delta
_{i}^{n}-2\rho(g^{-1}g^{\prime})_{i}^{n}+\rho^{2}(g^{-1}g^{\prime})_{i}^{k}
(g^{-1}g^{\prime})_{k}^{n} \cr &&
 -\rho g^{mk}\Gamma_{im}^{k}\nabla_{k}+\rho
tr(g^{-1}g^{\prime})\delta _{i}^{n}-\rho^{2}tr(g^{-1}g^{\prime})(g^{-1}g^{\prime})
_{i}^{n}-\rho g^{mk}\Gamma_{ik}^{n}\partial_{m} \cr&&
-\rho g^{km}(\partial_{m}\Gamma_{ik}^{n})-2\rho^{2}\partial_{\rho}(g^{-1}g^{\prime})_{i}^{n}\big]R_{nj}
+i \leftrightarrow j\bigg] \cr &&-\bigg[\big[\rho g^{km}(g_{im}-\rho g^{\prime}_{im})\nabla_{k}
+\rho g^{km}\partial_{m}(g_{ik}-\rho g^{\prime}_{ik})+\rho g^{km}(g_{ik}-\rho g^{\prime}_{ik})
\partial_{m}\big] \cr &&\times
\big[\nabla^{n}g^{\prime}_{nj}\nabla_{j}tr(g^{-1}g^{\prime})\big]
+i \leftrightarrow j\bigg] \cr&&-\bigg[\rho(g^{-1}g^{\prime})_{j}^{m}R_{im}+\rho (g^{-1}g^{\prime})_{i}^{n}
R_{nj}-\rho^{2}(g^{-1}g^{\prime})_{i}^{n}(g^{-1}g^{\prime})_{j}^{m}R_{nm}+i\leftrightarrow j\bigg].
\eea
Having these expressions and plugging them into the equations of motion one can write the equation we were
looking for.

\section{Linearized equations of motion for arbitrary $m$ }

In this appendix we find the linearized equations of motion for arbitrary $m$. To do this we need
to linearized different terms in the equations of motion as follows

\bea
\nabla^{2}R_{\rho\rho}&=&-2\rho^{2}\
tr(h^{4})-8\rho\ tr(h^{3})+tr(h^{\prime\prime})+\frac{1}{\rho}\
tr(h^{\prime})-\frac{1}{2}\rho\partial^{n}\partial_{n}\
tr(h^{\prime\prime})\cr
&&+\frac{1}{2\rho}R(h)+\partial^{i}\partial^{j}\ tr(h^{\prime}),\cr
\nabla^{2}R_{\rho i}&=&2\rho^2\partial^{m}h^{(3)}_{mi}-2\rho^{2}\partial_{i}\
tr(h^{(3)})+\frac{\rho}{2}\partial^{j}\partial_{j}[\partial^{m}h^{\prime}_{mi}-\partial_{i}\
tr(h^{\prime})]\cr
&&+\frac{1}{2}\partial_{i}\widetilde{R}(h)+4\rho\partial^{m}h^{\prime\prime}_{mi}-4\rho\partial_{i}\
tr(h^{\prime\prime})+2\partial_{i}\
tr(h^{\prime})-\partial^{m}h^{\prime}_{mi},\cr
\nabla^{2}R_{ij}&=&-8\rho^{3}h^{(4)}_{ij}-32\rho^2\;h^{(3)}_{ij}+4\rho^{(2)}\eta_{ij}
 tr(h^{(3)})+2\rho^{2}\eta_{ij}\widetilde{R}(h^{\prime\prime})\cr
&&-12\rho\ h^{\prime\prime}_{ij}-2\eta_{ij}\
 tr(h^{\prime})+4\rho\eta_{ij}\widetilde{R}(h^{\prime})+4\rho\eta_{ij}\
 tr(h^{\prime\prime}) \cr
&&+\frac{1}{2}\rho\eta_{ij}\partial^{m}\partial_{m}\widetilde{R}(h)+\rho\eta_{ij}\partial^{m}\partial_{m}
 tr(h^{\prime})-2\rho^{2}\partial^{m}\partial_{m}h^{\prime\prime}_{ij}
 \cr
&&-\eta_{ij}\widetilde{R}(h)-2\rho\partial_{i}\partial^{m}h^{\prime}_{mj}-2\rho\partial_{j}\partial^{m}h^{\prime}_{mi}+4\rho\partial_{i}\partial_{j}\
tr(h^{\prime}),
 \eea

\bea
\left(R^{\mu\nu}R_{\mu\nu}\right)^{(1)}&=&16\rho^{2}\
tr(h^{\prime\prime})-8\rho\ tr(h^{\prime})-8\
tr(h)-4\rho\widetilde{R},\cr
\left(R_{\rho}^{\sigma}R_{\sigma\rho}\right)^{(1)}&=&2\
tr(h^{\prime\prime}),\cr
\left(R_{i}^{\sigma}R_{\sigma\rho}\right)^{(1)}&=&-2\partial^{j}h^{\prime}_{ji}+2\partial_{i}
tr(h^{\prime}),\cr
\left(R_{i}^{\sigma}R_{\sigma j}\right)^{(1)}&=&8\rho
h^{\prime\prime}_{ij}+4\rho\eta_{ij}\
tr(h^{\prime})-2\eta_{ij}\widetilde{R}(h)+\frac{4}{\rho}h_{ij},
\eea

\bea
 E_{\rho\rho}&=&-\frac{1}{2}\ tr(h^{\prime\prime})
 +\frac{1}{2m^{2}}[4\rho^{2}\ tr(h^{(4)})+16\rho\
 tr(h^{(3)})-\frac{23}{2}\
tr(h^{\prime\prime})+\rho \partial^{i}\partial_{i}\
 tr(h^{\prime\prime})\cr&&+\frac{4}{\rho}\
 tr(h^{\prime})+2\partial^{i}\partial_{i}\
 tr(h^{\prime})+\frac{6}{\rho^{2}}\
 tr(h)+\frac{2}{\rho}
 R(h)-2\partial^{i}\partial^{j}h^{\prime}_{ij}]=0,\cr
E_{\rho i}&=&\frac{1}{2}(\partial^{m}h^{\prime}_{mi}-\partial_{i}\
tr(h^{\prime}))+\frac{1}{2m^{2}}[-4\rho^{2}\partial^{m}h^{(3)}_{mi}+4\rho^{2}\partial_{i}\
tr(h^{3})-\frac{1}{2}\partial^{m}h^{\prime}_{mi}-8\rho\partial^{m}h^{\prime\prime}_{mi}\cr
&&-\frac{3}{2}\partial_{i}\ tr(h^{\prime})
\rho\partial^{m}\partial_{m}\partial^{n}h^{\prime}_{ni}+\rho\partial^{m}\partial_{m}\partial_{i}\
tr(h^{\prime})-\partial_{i}\widetilde{R}(h)-8\rho\partial_{i}\
tr(h^{\prime\prime})]=0,\cr
E_{ij}&=&\frac{1}{2}\widetilde{R}(h)\eta_{ij}+\eta_{ij}\
tr(h^{\prime})-2\rho
h^{\prime\prime}_{ij}-\frac{1}{2\rho}h_{ij}+\frac{1}{2m^{2}} [34\rho
h^{\prime\prime}_{ij}+23\eta_{ij}\
tr(h^{\prime})+16\rho^{3}h^{(4)}_{ij} \cr &&+
64\rho^{2}h^{(3)}_{ij}-8\rho^{2}\eta_{ij}\
tr(h^{(3)})-4\rho^{(2)}\eta_{ij}\widetilde{R}(h^{\prime\prime})+\frac{23}{2}\eta_{ij}\widetilde{R}(h)
-56\rho\eta_{ij}\ tr(h^{\prime\prime}) \cr
&&-\rho\eta_{ij}\partial^{m}\partial_{m}\widetilde{R}(h)-2\rho\eta_{ij}\partial^{m}\partial_{m}\
tr(h^{\prime})+4\rho^2\partial^{m}\partial_{m}h^{\prime\prime}_{ij}
+4\rho\partial_{i}\partial^{m}h^{\prime}_{mj}+4\rho\partial_{j}\partial^{m}h^{\prime}_{mi}
\cr  &&+\frac{1}{2\rho}h_{ij}-8\rho\partial_{i}\partial_{j}\
tr(h^{\prime})-8\rho\partial_{i}\partial_{j}\
tr(h^{\prime})-8\rho\eta_{ij}\widetilde{R}(h^{\prime})+\frac{24}{\rho}\eta_{ij}\
tr(h) ]=0
\eea
Putting these expressions together one can find the linearized equations of motion.

\section{Linearized action up to quadratic terms }

To find the on-shell action we need to plug the classical solution of the equations of motion to
the complete action. Of course since in our case where $m^2=1/2$ we do not need the Gibbons-Hawking boundary
term, the whole contributions  to the on-shell action come from the boundary terms given in
\eqref{AA}. Since we are only interested in the two point functions we will need to know the action
up to the quadratic terms. In what follows we will write the contribution of each term to the on-shell action.

\bea
\int_{\partial R}d^2x\sqrt{-\gamma}\
n_{\mu}\left[
G^{\alpha\beta}\delta\Gamma^{\mu}_{\alpha\beta}-G^{\alpha\mu}\delta\Gamma^{\beta}_{\alpha\beta}\right]=\frac{1}{2}\int_{\partial R}d^2x
[\frac{3}{\rho}h^{ij}]h_{ij}+[-3h^{ij}]h^{\prime}_{ij}
\eea

\bea
\int_{\partial R} d^2x \sqrt{-\gamma}\;n_{\mu}
\left[(2R^{\alpha\beta}-\frac{3}{4}RG^{\alpha\beta})\
\delta\Gamma_{\alpha\beta}^{\mu} \right]
&=&\frac{1}{2}\int_{\partial R}d^2x \
[-\eta^{ij}\widetilde{R}(h)-2\eta^{ij} tr(h^{\prime})+4\rho
h^{\prime\prime ij}\cr &&\;\;\;\;\;\;\;\;\;\;\;\;\;\;\;\;\;\;+\frac{1}{2\rho}h^{ij}]h_{ij}+[\rho
\eta^{ij}\widetilde{R}(h)-\frac{1}{2}h^{ij}\cr &&\;\;\;\;\;\;\;\;\;\;\;\;\;\;\;\;\;\;+2\rho \eta^{ij}
tr(h^{\prime})-4\rho^{2}h^{\prime\prime\
ij}]h^{\prime}_{ij},
\eea

\bea
\int_{\partial R} d^2x \sqrt{-\gamma}\;n_{\mu}
\left[(2R^{\alpha\mu}-\frac{3}{4}RG^{\alpha\mu})\
\delta\Gamma_{\alpha\beta}^{\beta} \right]
&=&\frac{1}{2}\int_{\partial R}d^2x \
[-\frac{1}{\rho}h^{ij}]h_{ij}+[4\rho^{2}\eta^{ij}\
tr(h^{\prime\prime})+h^{ij}]h^{\prime}_{ij} \cr
&&\;\;\;\;\;\;\;\;\;\;\;\;\;\;\;\;\;-\rho\eta^{in}\eta^{jm}[\partial^{k}h^{\prime}_{kn}-\partial_{n}
tr(h^{\prime})]\partial_{i}h_{mj}, \cr &&
\eea

\bea
\int_{\partial R} d^2x \sqrt{-\gamma}\;n_{\beta}
\left[(\nabla _{\mu} R^{\alpha\beta}G^{\mu\nu})\ \delta
G_{\alpha\nu} \right]& =&\frac{1}{2}\int_{\partial R}d^2x \bigg[-\eta^{ij}\widetilde{R}(h)-\rho\eta^{im}\partial^{j}(\partial^{n}h^{\prime}_{nm}-\partial_{m}
tr(h^{\prime}))\cr &&\;\;\;\;\;\;\;\;\;\;\;\;\;\;\;\;\;-2\rho\eta^{ij} tr(h^{\prime\prime})-\eta^{ij}
tr(h^{\prime})+2\rho h^{\prime\prime\ ij}\bigg]h_{ij},\cr &&
\eea

\bea
\int_{\partial R} d^2x \sqrt{-\gamma}\;n_{\beta}
\left[(\nabla _{\mu} R^{\alpha\nu}G^{\mu\beta})\ \delta
G_{\alpha\nu} \right] & =&\frac{1}{2}\int_{\partial R}d^2x
\bigg[-\eta^{ij}\widetilde{R}(h)-2\eta^{ij} tr(h^{\prime})+8\rho
h^{\prime\prime\
ij}\cr &&\;\;\;\;\;\;\;\;-\rho\eta^{ij}\widetilde{R}(h^{\prime})-2\rho\eta^{ij}
tr(h^{\prime\prime})+4\rho^{2}h^{\prime\prime\prime\;ij}\bigg]h_{ij}.\cr &&
 \eea
Altogether  we find

\bea
 S_{(2)}&=& \frac{1}{32\pi G}\int_{\partial R}
d^2x\;[(\frac{3}{\rho})h^{ij}h_{ij}+(-3)h^{ij}h^{\prime}_{ij}]
 \cr &&-\frac{1}{m^{2}} \frac{1}{32\pi G}\int_{\partial R} d^2x
\bigg[(\frac{3}{2\rho})h^{ij}h_{ij}+(\frac{-3}{2})h^{ij}h^{\prime}_{ij}+8\rho
h^{\prime\prime ij}h_{ij}+4\rho^{2}h^{\prime\prime\prime\;
ij}h_{ij}  \cr
&&-\eta^{ij}\widetilde{R}(h)h_{ij}+2\rho\eta^{ij}tr(h^{\prime\prime})h_{ij}
-\rho\eta^{ij}\widetilde{R}(h^{\prime})h_{ij}-2\eta^{ij}
tr(h^{\prime})h_{ij}  \cr &&
-4\rho^{2}\eta^{ij}tr(h^{\prime\prime})h^{\prime}_{ij}+\rho\eta^{ij}\widetilde{R}(h)h^{\prime}_{ij}
+2\rho\eta^{ij}tr(h^{\prime})h^{\prime}_{ij}-4\rho^{2}h^{\prime\prime
ij}h^{\prime}_{ij}  \cr
&&+2\rho\eta^{im}\partial^{j}[\partial^{n}h^{\prime}_{mn}-\partial_{m}
tr(h^{\prime})]h_{ij}+\rho\eta^{in}\eta^{jm}[\partial^{k}h^{\prime}_{kn}-\partial_{n}
tr(h^{\prime})]\partial_{i}h_{mj}\bigg]. \eea Setting $m^2=1/2$
and using the results of the section (3.2) one find the quadratic
on-shell action \eqref{second.action}.

Note that to write the above expression for arbitrary $m$ we had
to consider the contribution of the Gibbons-Hawking term and
another counterterm. To do this we need the form of the
Gibbons-Hawking term in the Fefferman-Graham coordinates which is
\bea S_{(G.H)}=\int_{\partial
R}d^{2}x\sqrt{-g}\big(\frac{1}{\rho}g_{ij}-g^{\prime}_{ij}\big)g^{ij}.
\eea
Therefore the variation of the Gibbons-Hawking term reads

\bea \delta I_{(G.H)}=\int_{\partial
R}d^{2}x\sqrt{-g}\left[\left(\frac{1}{\rho}g_{ij}-g^{\prime}_{ij}
+\frac{1}{2} tr(g^{-1}g^{\prime})g_{ij}\right)\delta
g^{ij}-(g^{ij})\delta g^{\prime}_{ij}\right]. \eea

Note that to cancel an infinity arising from a term like
$(\frac{1}{\rho}h^{ij}h_{ij})$ we need another counterterm given by
\bea
S_{c.t.2}=\frac{1}{8\pi G}\int d^{2}x \sqrt{\gamma}\;
\big(-1+\frac{1}{4}R[\gamma]\log(\rho_{0})\big).
\eea
The variation of this counterterm in Fefferman-Graham coordinates is
\bea
\delta S_{c.t.2}=-\frac{1}{16\pi G}\int d^{2}x \sqrt{-g}\;
\big[\frac{1}{\rho}\;g^{ij}\delta g_{ij}\big].
\eea

\section{Notations}

In this appendix we present the notations used in section 4.2. Let us define the new coordinates as follows
\bea
  z=x-t=x^{1}-x^{2},\;\;\;\;\;\;\;\;\;\;
  \bar{z}=x+t=x^{1}+x^{2}.
\eea
Therefore
\bea
\partial^{1}=\frac{\partial x^{1}}{\partial
z}\;\partial^{z}+ \frac{\partial x^{1}}{\partial
\bar{z}}\;\partial^{\bar{z}}
=\frac{1}{2}(\partial^{z}+\partial^{\bar{z}}),\;\;\;\;\;\;
\partial^{2}=\frac{\partial x^{2}}{\partial z}\;\partial^{z}+
\frac{\partial x^{2}}{\partial
\bar{z}}\;\partial^{\bar{z}}
  =\frac{1}{2}(-\partial^{z}+\partial^{\bar{z}})
\eea
In this notation one has
\bea\eta_{ij}=\left(
             \begin{array}{cc}
               1 & 0 \\
               0 & -1 \\
             \end{array}
           \right)\;\;\;\;\;\;\;\;\;\;\eta_{z}^{ij}=\left(
                                                          \begin{array}{cc}
                                                            0 & 2 \\
                                                            2 & 0 \\
                                                          \end{array}
                                                        \right).
\eea
Using the fact that

\bea
\partial_{1}=\eta_{11}\partial^{1}\;\;\;\;\;\partial_{2}=
\eta_{22}\partial^{2},\;\;\;\;\;
\partial^{z}=\eta^{z\bar{z}}\partial_{\bar{z}}\;\;\;\;\;
\partial^{\bar{z}}=\eta^{\bar{z}z}\partial_{z}
\eea
one gets
\bea
\partial^{1}\partial_{1}+\partial^{2}\partial_{2}=\partial^{z}
\partial^{\bar{z}}=4\partial_{z}\partial_{\bar{z}}
\eea
Using this notations, for example, we have
\bea
&&b_{(2)11}=b_{(2)zz}+b_{(2)\bar{z}\;\bar{z}}+b_{(2)z\bar{z}}+b_{(2)\bar{z}z}\cr
&&b_{(2)22}=b_{(2)zz}+b_{(2)\bar{z}\;\bar{z}}-b_{(2)z\bar{z}}-b_{(2)\bar{z}z}\cr
&&b_{(2)12}=-b_{(2)zz}+b_{(2)\bar{z}\;\bar{z}}+b_{(2)z\bar{z}}-b_{(2)\bar{z}z}\cr
&&b_{(2)21}=-b_{(2)zz}+b_{(2)\bar{z}\;\bar{z}}-b_{(2)z\bar{z}}+b_{(2)\bar{z}z}
\eea

As an example let us recast the one point functions in terms of $z\bar{z}$ coordinates.
To proceed we note that the second order expansion of action \eqref{second.action.final} in
these coordinates is given by
\bea
S_{2,tot}&=&\frac{1}{16\pi G}\;\frac{1}{2}\int_{\partial R}dz d\bar{z}\bigg[
32\rho h^{\prime\prime}_{zz}h_{\bar{z}\;\bar{z}}+32\rho h^{\prime\prime}
_{\bar{z}\;\bar{z}}h_{zz}+16\rho^{2} h^{\prime\prime\prime}_{zz}
h_{\bar{z}\;\bar{z}}+16\rho^{2} h^{\prime\prime\prime}_{\bar{z}\;\bar{z}}
h_{zz}\\ &&-16\rho(\partial\bar{\partial}h^{\prime}_{zz})h_{\bar{z}\;\bar{z}}
-16\rho(\partial\bar{\partial}h^{\prime}_{\bar{z}\;\bar{z}})h_{zz}
-16\rho^{2}h^{\prime\prime}_{zz}h^{\prime}_{\bar{z}\;\bar{z}}
-16\rho^{2}h^{\prime\prime}_{\bar{z}\;\bar{z}}h^{\prime}_{zz}
+16\rho h^{\prime}_{zz}h^{\prime}_{\bar{z}\;\bar{z}}\bigg].\nonumber
\eea
Using this expression the one point function of energy-momentum tensor, $T$, reads\footnote{Note that $\sqrt{-g_{(0)}}=\frac{1}{2}.$}
\bea
\langle T_{zz}\rangle &=&\frac{4\pi}{\sqrt{-g_{(0)}}}\frac{\delta S_{2,tot}}{\delta h^{zz}}=
\frac{4\pi}{\sqrt{-g_{(0)}}}(-\frac{1}{4})\frac{\delta S_{2,tot}}{\delta h_{\bar{z}\;\bar{z}}}
\\ &=& {4\pi}(\frac{1}{\frac{1}{2}})(\frac{1}{16\pi G\times 2})(\frac{-1}{4})\big[16 b_{(2)\;zz}-16\partial
\bar{\partial}b_{(0)\;zz}\big]=\frac{1}{4 G}\big[-4b_{(2)\;zz}+4\partial\bar{\partial}
b_{(0)\;zz}\big].\nonumber
\eea
One the other hand for $\langle t_{zz}\rangle$ we first need to calculate
\bea
&&\frac{1}{\rho}\frac{\delta S_{2,tot}}{\delta h^{\prime\;zz}}=\frac{1}{\rho}
(\frac{-1}{4})\frac{\delta S_{2,tot}}{\delta h^{\prime}_{\bar{z}\;\bar{z}}}
\\ &&=(\frac{1}{16\pi G\times 2})\big[-4b_{(2)zz}\log(\rho)-4g_{(2)_{zz}}
+4(\partial\bar{\partial} b_{(0)zz})\log(\rho)+4\partial\bar{\partial}
g_{(0)zz}+\rho-term\big].\nonumber
\eea
As a result we arrive at
\bea
\langle t_{zz}\rangle &=&\frac{4\pi}{\sqrt{-g_{(0)}}}\frac{1}{\rho}\frac{\delta S_{2,tot}}
{\delta h^{\prime\;zz}}-\log(\rho)\frac{4\pi}{\sqrt{-g_{(0)}}}\frac{\delta S_{2,tot}}
{\delta h^{zz}}\\ &=&(4\pi)\frac{1}{\frac{1}{2}}(\frac{1}{16\pi G\times 2})\big[
 -4g_{(2)\;zz}+4\partial\bar{\partial}g_{(0)zz}\big]=
\frac{1}{4 G}\big[
-4g_{(2)\;zz}+4\partial\bar{\partial}g_{(0)zz}\big].\nonumber
\eea




\end{document}